\documentclass[a4paper,11pt]{article}
\usepackage{amsmath,amssymb,amsthm,amsxtra,overpic,bbm,bm,epsfig,ulem,multirow}
\usepackage{color,cite}
\usepackage{epstopdf}
\usepackage{booktabs}
\usepackage{footnote}
\usepackage{rotating}
\usepackage{color,cite}

\usepackage[affil-it]{authblk}

\usepackage{lineno}

\def\thefootnote{\fnsymbol{footnote}}

\usepackage{array}

\title{Prospects for Detecting the Diffuse Supernova Neutrino Background with JUNO}

\author[6,5]{Angel Abusleme}
\author[46]{Thomas Adam}
\author[67]{Shakeel Ahmad}
\author[67]{Rizwan Ahmed}
\author[56]{Sebastiano Aiello}
\author[67]{Muhammad Akram}
\author[30]{Fengpeng An}
\author[23]{Qi An}
\author[56]{Giuseppe Andronico}
\author[68]{Nikolay Anfimov}
% ...
\affil[5]{Millennium Institute for SubAtomic Physics at the High-energy Frontier (SAPHIR), ANID, Chile}
\affil[6]{Pontificia Universidad Cat\'{o}lica de Chile, Santiago, Chile}
% ...
\affil[23]{University of Science and Technology of China, Hefei, China}
% ...
\affil[30]{East China University of Science and Technology, Shanghai, China}
% ...
\affil[46]{IPHC, Universit\'{e} de Strasbourg, CNRS/IN2P3, F-67037 Strasbourg, France}
% ...
\affil[56]{INFN Catania and Dipartimento di Fisica e Astronomia dell Universit\`{a} di Catania, Catania, Italy}
% ...
\affil[67]{Pakistan Institute of Nuclear Science and Technology, Islamabad, Pakistan}
\affil[68]{Joint Institute for Nuclear Research, Dubna, Russia}
% ...

\begin{document}

\maketitle






\newpage

\begin{abstract}
We present the detection potential for the diffuse supernova neutrino background (DSNB) at the Jiangmen Underground Neutrino Observatory (JUNO),
using the inverse-beta-decay (IBD) detection channel on free protons.
We employ the latest information on the DSNB flux predictions, and investigate in detail the background and its reduction for the DSNB search at JUNO.
The atmospheric neutrino induced neutral current (NC) background turns out to be the most critical background, whose uncertainty is carefully evaluated
from both the spread of model predictions and an envisaged \textit{in situ} measurement.
We also make a careful study on the background suppression with the pulse shape discrimination (PSD) and triple coincidence (TC) cuts.
With latest DSNB signal predictions, more realistic background evaluation and PSD efficiency optimization, and additional TC cut,
JUNO can reach the significance of 3$\sigma$ for 3 years of data taking, and achieve better than 5$\sigma$ after 10 years for a reference DSNB model. 
In the pessimistic scenario of non-observation, JUNO would strongly improve the limits and exclude a significant region of the model parameter space.
\end{abstract}

\begin{flushleft}
\hspace{0.8cm} Keywords: diffuse supernova neutrino background, detection potential, JUNO
\end{flushleft}

\def\thefootnote{\arabic{footnote}}
\setcounter{footnote}{0}

\newpage

\section{Introduction}

The explosion of the massive core-collapse supernova (SN) is one of the most powerful astrophysical phenomena in the Universe, which can release around $10^{53}$ ergs of energy, among which 99\% is in the form of neutrinos and antineutrinos. Galactic SNe are rather rare~\cite{Rozwadowska:2021lll}, and thus the chance of a detection during the lifetime of an experiment is low. The diffuse supernova neutrino background (DSNB), which is the integrated neutrino signal from all the SN explosions in the Universe, is expected to be visible in large underground neutrino detectors. The detection of DSNB signals is important for the cosmology. It holds the precise information on the average core-collapse SN neutrino spectrum, the cosmic star-formation rate and the fraction of failed black-hole forming SNe~\cite{Ando:2004hc, Beacom:2010kk, Lunardini:2010ab}.

The existing and future large water-Cherenkov and liquid-scintillator (LS) detectors have promising potential to first observe the DSNB via the inverse-beta-decay (IBD) reaction, $\overline{\nu}^{}_e + p \to e^+ + n$, which consists of a prompt signal of positron and a delayed signal of neutron capture {on Hydrogen or Gadolinium}.
Super-Kamiokande (SK) has searched for the DSNB~\cite{Malek:2002ns, Bays:2011si,Super-Kamiokande:2021jaq,Zhang:2013tua}, but no signal has been found yet.
The new Gadolinium-doped SK (SK-Gd) will greatly improve the neutron tagging efficiency and hence significantly reduce the background level, increasing the sensitivity of the DSNB~\cite{Beacom:2003nk, Watanabe:2008ru, Horiuchi:2008jz, Labarga:2018fgu}.
Compared to the water-Cherenkov detectors, the LS detectors have lower energy thresholds, higher energy resolution, and more than 98\% neutron tagging capability{\cite{JUNO:2022mxj}}. The DSNB search in LS detectors has been previously taken up by KamLAND~\cite{KamLAND:2021gvi,Collaboration:2011jza} and Borexino~\cite{Borexino:2019wln}, whose observation of the IBD-like signal in the selected energy range is highly consistent with the expected background, setting the upper limits on the total DSNB flux.

The Jiangmen Underground Neutrino Observatory (JUNO)~\cite{An:2015jdp}, which is under construction in South China and will be online in 2023, would be the largest ever LS detector.
In this work, we give a comprehensive study on the prospects for detecting the DSNB signal at JUNO, updating the results obtained in 2015 in Ref.~\cite{An:2015jdp}.
Firstly, we revisit the DSNB signal predictions at JUNO based on the latest properties of large-scale SN numerical simulation.
Then relevant background budgets will be investigated. The dominant one is from the neutral-current (NC) interaction of atmospheric neutrinos with
$^{12}$C nuclei, which surpasses the DSNB signal by more than one order of magnitude.
The systematic uncertainty of the NC background is evaluated from both the spread of model predictions and an envisaged \textit{in situ} measurement.
We provide a detailed evaluation of the efficiencies of the pulse-shape discrimination (PSD) technique and the triple-coincidence (TC) cut for the NC background.
We find that the prospects for detecting the DSNB signal at JUNO are promising. For a reference DSNB flux model, the significance can reach the level of 3$\sigma$ for around 3 years of data taking, and better than 5$\sigma$ after 10 years.
A non-observation would strongly improve the limits of the DSNB flux and exclude a significant region of the DSNB model space.

This paper is organized as follows. In Sec.~2 we give a brief introduction of the JUNO detector.
Then we present the DSNB signal prediction in Sec.~3 and the background budget evaluations in Sec.~4.
Sec.~5 is devoted to the background suppression techniques, including the PSD and TC cuts.
Finally, we present the sensitivity study of the DSNB signal in Sec.~6 and conclude with a few remarks in Sec.~7.

\section{JUNO Detector}
\label{sec:junodect}

The JUNO experiment is located at Jiangmen in Guangdong province, China at equal distance from the Taishan and Yangjiang nuclear power plants,
with the primary goals of determining the neutrino mass ordering~\cite{An:2015jdp,Li:2013zyd,JUNO:2021vlw} and the precision measurement of oscillation parameters~\cite{JUNO:2022mxj} with reactor antineutrinos, together with 
other physics program, including studies of neutrinos from
the Sun~\cite{JUNO:2020hqc}, the planet Earth~\cite{Han:2015roa}, the atmosphere~\cite{JUNO:2021tll}, and the core collapse SNe~\cite{Lu:2016ipr} as well as the exploration
of physics beyond the Standard Model~\cite{An:2015jdp}.

\begin{figure}
	\centering
	\includegraphics[scale=0.4]{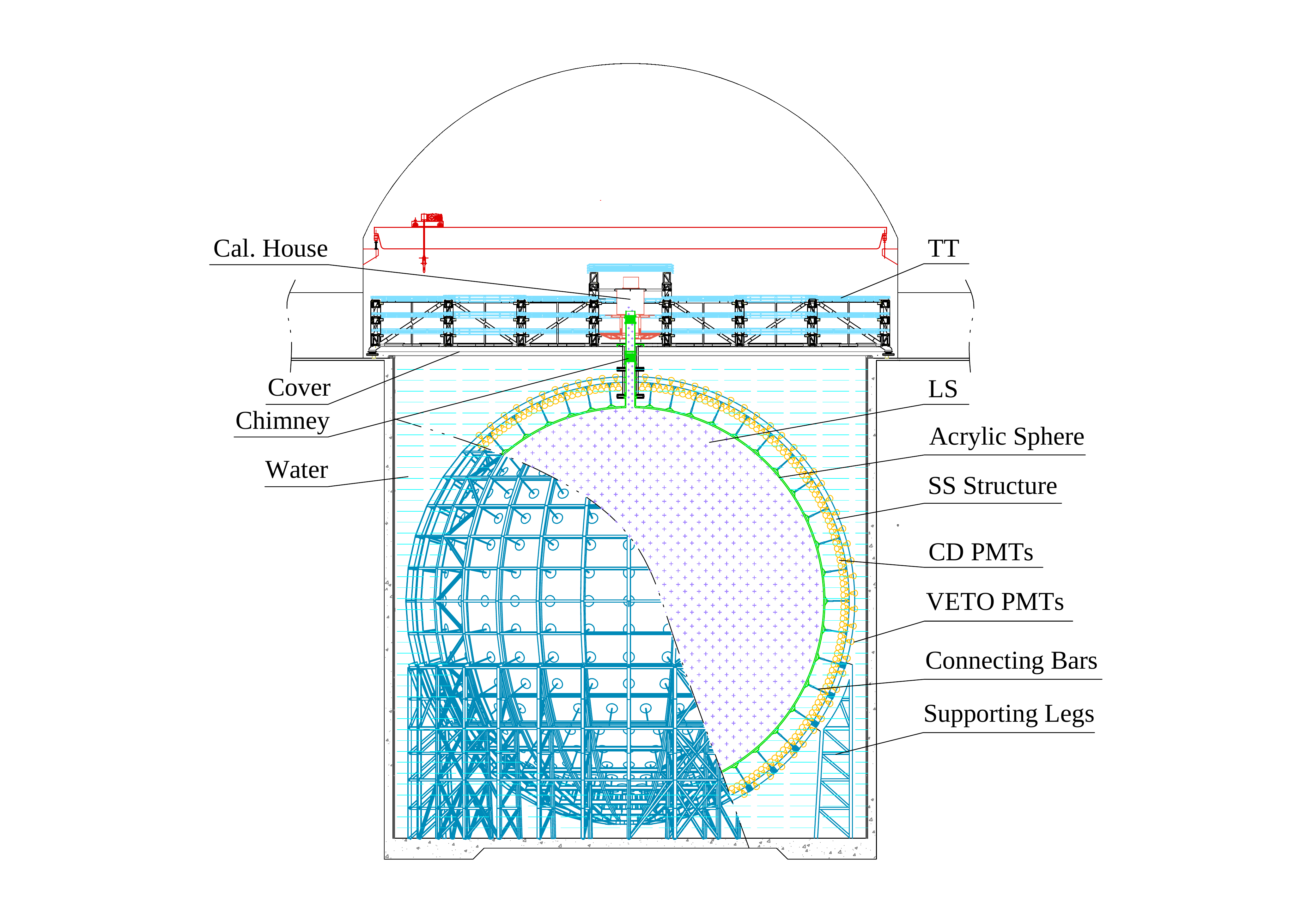}
	\setlength{\abovecaptionskip}{-0.5cm}
	\caption{\label{fig:detector} Schematic of the JUNO detector complex, which is composed of the Central Detector (CD), water buffer, and the veto detectors. See the text for more details.
	}
\end{figure}

JUNO is an underground LS detector of 20 kton with an overburden of around 700 meter of rock (1800 meter water equivalent) for shielding of the cosmic rays. This results in a muon rate of 0.004 $\rm Hz/m^2$ and an average muon energy of 207 GeV at the detector location.
The JUNO detector complex is composed of the Central Detector (CD), water buffer, and the veto detectors,
which are illustrated in Fig.~\ref{fig:detector}.
From the inner to outer layers, the CD contains 20 kton LS filled in an acrylic shell with an inner diameter of 35.4 m.
It is immersed in a cylindrical water pool with the diameter and height of both 43.5 m. There are 17,612 high-quantum-efficiency 20-inch photomultipliers (PMTs) closely packed around the LS sphere. Another 25,600 3-inch PMTs are installed in the gaps between the 20-inch PMTs to further improve the neutrino energy measurements. The
water buffer is filled between the acrylic shell and PMTs supported with the stainless steel structure.
The Veto System is designed to tag muons with high efficiency and precisely reconstruct their tracks for effective background reduction.
The Veto System includes the water-Cherenkov detector surrounding the CD to shield the neutrons and the natural radioactivity from the rock and the Top Tracker.
The water-Cherenkov detector contains 35 kton ultrapure water,
which is supplied and maintained by a circulation system. The Cherenkov light is detected by ~2400 20-inch PMTs. 
{Its muon detection efficiency is expected to be greater than 99$\%$}.
The Top Tracker is made from the reused plastic scintillator from the OPERA experiment. It covers half of the water pool on the top with a 3-layers configuration.
Each detector module is read out at both ends by multi-anode PMTs.

\section{DSNB signal prediction}\label{sec:dsnb}

The DSNB signal calculation depends on a variety of important ingredients~\cite{Priya:2017bmm,Kresse:2020nto,Horiuchi:2020jnc}. The first one is the cosmological SN rate as a function of the progenitor mass and redshift, which is the link to the cosmic history of the star formation. The second ingredient is the average energy spectrum of SN neutrinos. According to the latest large-scale SN numerical simulation~\cite{Kresse:2020nto,Horiuchi:2020jnc}, there are more astrophysical or physical effects which may alter the DSNB signals, including the fraction of failed black-hole-forming SNe~\cite{Priya:2017bmm,Kresse:2020nto} and binary interactions~\cite{Horiuchi:2020jnc}. In this paper we shall consider the contribution of the failed SNe, which will feature a hotter neutrino energy spectrum compared to the neutron-star-forming SNe (i.e., successful SNe), {and they could represent a fraction of all the SNe ranging from around 20\% to 40\%~\cite{Kresse:2020nto}.}

The isotropic DSNB flux is obtained by an integration of the cosmic redshift $z$ by 
\begin{equation}
\label{DSNB}
	\frac{d \phi}{d E_\nu}=\int_{0}^{z_{\rm max}} R_{\mathrm{SN}}(z) \frac{d N\left(E_\nu^{\prime}\right)}{d E_\nu^{\prime}}(1+z)\left|\frac{c d t}{d z}\right| d z,
\end{equation} 
where ${c}$ is the speed of light, $z_{\rm max}$ is the maximal redshift boundary to be covered in the integration, ${|{dt}/{dz}}|^{-1} = H_{0}(1+z) [\Omega_{\Lambda}+\Omega_{m}(1+z)^{3}]^{\frac{1}{2}}$ includes the present-day Hubble constant
(${H_{0}\simeq67.4 \,\rm km\cdot s^{-1}\cdot Mpc^{-1}}$~\cite{ParticleDataGroup:2020ssz}), the ratios of the energy density of matter and the cosmological constant ($ {\Omega_{m}}\simeq0.3$ and ${\Omega_{\Lambda}}\simeq0.7$).
Due to the redshift, a neutrino detected at the energy ${E_{\nu}}$ was emitted at a higher energy $E_\nu^{\prime}={E_{\nu}(1+z)}$.

In the DSNB flux, $ {{dN}/{dE_{\nu}}}$ is the average SN neutrino spectrum, which has contributions from both successful and failed SNe:
\begin{equation}\label{Flux_CCSN}
\frac{dN(E_\nu)}{dE_\nu} = (1-f_\mathrm{BH})\frac{dN_\mathrm{SN}(E_\nu)}{dE_\nu}
+ f_\mathrm{BH} \frac{dN_\mathrm{BH}(E_\nu)}{dE_\nu},
\end{equation}
where $f_\mathrm{BH}$ is the fraction of the failed SNe where we take a reference value of $ f_\mathrm{BH} = 27\% $~\cite{Priya:2017bmm} and scan a reasonable region from 0 to 40\% for the sensitivity study.

The average energy spectrum for both successful and failed SNe is given as~\cite{Priya:2017bmm}
\begin{equation}\label{Spectrum_CCSN}
\frac{dN_\nu}{dE_\nu} = \frac{E_\mathrm{total}}{\langle E_\nu \rangle^2}
\frac{(1+\gamma_\alpha)^{1+\gamma_\alpha}}{\Gamma(1+\gamma_\alpha)}
\left( \frac{E_\nu}{\langle E_\nu \rangle} \right)^{\gamma_\alpha}
 \exp\left( -(1+\gamma_\alpha) \frac{E_\nu}{\langle E_\nu \rangle} \right),
\end{equation}		
where $ E_\mathrm{total} $ is the total energy and the $ \langle E_\nu \rangle $ is the average energy of the SN neutrinos, and the spectral index
\begin{equation}\label{gamma_alpha}
\gamma_\alpha = \frac{\langle  E_\nu^2 \rangle - 2 \langle  E_\nu \rangle^2 }
{\langle  E_\nu \rangle^2 -\langle  E_\nu^2 \rangle}\,.
\end{equation}
For the failed SNe, we follow the model described in Ref.~\cite{Priya:2017bmm} and assume the model parameters of $E_\mathrm{total} = 8.6\times 10^{52} \ \rm{erg} $, $ \langle E_\nu \rangle = 18.72 \ \rm{MeV} $ and $ \langle  E_\nu^2 \rangle = 470.76\ \mathrm{MeV}^2$. Meanwhile, for the successful SNe, we take the reference value of $ E_\mathrm{total} = 5.0\times 10^{52} \ \mathrm{erg} $, $\gamma_\alpha = 3$ and $\langle  E_\nu \rangle = 15\ \mathrm{MeV} $, but scan a range of $\langle  E_\nu \rangle$ from 12 to 18 MeV in the sensitivity study. Notice that in general the failed SNe will have relatively larger average energies and thus hotter neutrino energy spectrum compared to the successful SNe.

$ R_\mathrm{SN}(z)$ is the SN rate at the redshift $z$, which can be derived from the star formation (SF) rate.
$ R_\mathrm{SF}$, which can be written as~\cite{Salpeter:1955it,Hopkins:2006bw}:
\begin{equation}\label{R_CCSN}
R_\mathrm{SN}(z) = R_\mathrm{SF}(z)
\frac{\int_{8}^{125} \psi(M) dM }{\int_{0.1}^{125}M \psi(M) dM}\,,
\end{equation}
where $M$ is the stellar mass in the unit of solar mass, [0.1, 125] and [8, 125] are the mass integration ranges of all the stars and those undergo core collapse SN explosions, respectively.
$\psi(M) \propto M^{-2.35} $ is the initial mass function (IMF)\cite{Salpeter:1955it}.
In the current study, we employ the relative redshift dependence as
\begin{equation}\label{R_SF}
R_\mathrm{SF}(z) \propto \frac{(a+bz)h}{1+(z/c)^d}\,,
\end{equation}
which is an empirical parametrization based on astrophysical observations, 
with the best fit values of $a=0.0170$, $b=0.13$, $c=3.3$, $d=5.3$ and $h = 0.7$~\cite{Hopkins:2006bw}.
A reference value of the absolute present SN rate at $z=0$ is taken as $R_\mathrm{SN}(0) = 1.0 \times 10^{-4}\,\mathrm{yr^{-1}\,Mpc^{-3}} $~\cite{Horiuchi:2008jz}.
However, due to many astrophysical factors the SN rate is still uncertain, we take a wide range of $R_\mathrm{SN}(0)$ varying the reference value by a factor of two, i.e., $0.5 \times 10^{-4}\,\mathrm{yr^{-1}\,Mpc^{-3}}\leq R_\mathrm{SN}(0)\leq 2.0 \times 10^{-4}\,\mathrm{yr^{-1}\,Mpc^{-3}} $.

Finally, in order to calculate the observed DSNB energy spectrum at JUNO, we need to consider the IBD cross section, the target mass and detector response.
{We take the free proton number in the JUNO LS as ${7.15\times 10^{31} \, \mathrm{kton^{-1}}}$~\cite{An:2015jdp}, whose mass fraction is around 12$\%$}. The differential IBD cross section is taken from Ref.~\cite{Strumia:2003zx}, and an energy resolution of 3\% is assumed~\cite{An:2015jdp}.

\section{Background evaluation}
\label{sec:bkgcal}

In this section, we turn to the background calculation relevant for the DSNB search at JUNO. First, there are two important IBD backgrounds from other $\overline{\nu}^{}_e$ sources. In the vicinity of the low-energy part of the DSNB $\overline{\nu}^{}_e$ spectrum, an irreducible background originates from $\overline{\nu}^{}_e$'s emitted from nearby nuclear reactors, whose fluxes are highly suppressed above the neutrino energy of around $\mathcal{O}(10)$ MeV. The high-energy part of the indistinguishable background is composed of the IBD interactions of the atmospheric $\overline{\nu}^{}_e$, which gradually increases as the neutrino energy grows. Therefore, the optimal energy window for the DSNB is between these two backgrounds.

Second, there are also non-IBD backgrounds from the cosmic muon spallation process. It can be well controlled by proper muon veto strategies. The fast neutron (FN) background is generated by muon spallation in the rock surrounding the detector. The event rate is higher at the surface of the CD, and can be effectively reduced by a fiducial volume cut. 
When energetic cosmic muons travel through the LS, they can interact with $^{12}$C nuclei and produce radioactive isotopes, among which the $\beta$-n decays of $^{9}$Li and $^{8}$He can mimic the $\overline{\nu}^{}_e$ IBD reaction, which is called the $^{9}$Li/$^{8}$He background.

Finally we have to face the non-IBD background induced by atmospheric neutrino interactions with the $^{12}$C nuclei. When high energy atmospheric neutrinos interact with the ${^{12}{\rm C}}$ nuclei via the charged-current (CC) or neutral-current (NC) interaction channel, copious neutrons, protons, $\gamma$'s and $\alpha$'s are generated together with the associated leptons, where the interactions with one single neutron capture may contaminate the IBD signals. {The CC background on $^{12}$C is usually accompanied by a high energy charged lepton, whose prompt energies are relatively higher and can be removed by a proper selection of the signal energy window.}
The most critical background is the NC background, which has been carefully studied in a general way in Refs.~\cite{Cheng:2020aaw,Cheng:2020oko}, and is estimated to be one order of magnitude higher than the typical DSNB signal.

\subsection{Reactor $\overline{\nu}^{}_e$}\label{sec:bkgrea}

Reactor $\overline{\nu}^{}_e$'s are emitted from the $\beta$-decays of neutron-rich fission fragments, mainly from four fission isotopes, $^{235}$U, $^{238}$U, $^{239}$Pu and $^{241}$Pu. Here we consider eight reactors from the Yangjiang and Taishan nuclear power plants, with a total thermal power of 26.6 $\rm GW_{th}$, and an average baseline of around 52.5 km. Contributions from other reactors are sub-dominant and neglected in the current study. 
Our calculation of the IBD rate and spectrum from reactor $\overline{\nu}^{}_e$'s follows the description in Ref.~\cite{An:2015jdp}. The IBD rate with the oscillation effect is expected at 1514.8 $\rm {yr}^{-1}{kt}^{-1}$. The spectral shape is taken from {the Huber-Muller model~\cite{Mueller:2011nm,Huber:2011wv}}, with the energy range up to 12 MeV. Since the yield and spectrum of high energy reactor $\overline{\nu}^{}_e$'s are rather uncertain, we currently take the low energy threshold as 12 MeV and neglect the background from reactor $\overline{\nu}^{}_e$'s. Lowering this threshold shows negligible effects on the DSNB sensitivity.

%

\subsection{Atmospheric $\overline{\nu}^{}_e$}\label{sec:bkgccibd}

Atmospheric $\overline{\nu}^{}_e$'s below 100 MeV can also induce the IBD signals. The atmospheric neutrino flux at low energies has
been calculated by several different groups from Battistoni {\it et al.}~\cite{Battistoni:2005pd}, Gaisser {\it et al.}~\cite{Gaisser:1988ar} and Honda {\it et al.}~\cite{Honda:1990sx} for the location of SK, showing significant model variations. A recent study calculated the
new atmospheric neutrino flux from stopped muons in the Earth~\cite{Guo:2018sno}. We employ a new calculation of low energy fluxes from the Honda group for the JUNO site~\cite{Hondahomepage}, and assume a systematic uncertainty of 50\% to cover the large flux variations for neutrino energies below 100 MeV. The rate and energy spectrum of the atmospheric neutrino induced IBD signal can be calculated in the same way as the DSNB and reactor $\overline{\nu}^{}_e$'s.

%

\subsection{Cosmogenic $^{9}$Li/$^{8}$He}\label{sec:bkgli9}


The cosmogenic production rates of $^{9}$Li and $^{8}$He have been measured in KamLAND~\cite{Abe:2009aa} and Borexino~\cite{Bellini:2013pxa}.
The yield of the radioactive isotopes $^{9}$Li and $^{8}$He is proportional to $R_{\mu}\cdot E^{0.74}_{\mu}$~\cite{Wang:2001fq} where $R_{\mu}$ is the muon rate and $E_{\mu}$ is the average muon energy at the detector. The $^{9}$Li/$^{8}$He yield is also related to the LS density and the average path length of muons in the LS. Our calculations of the $^{9}$Li and $^{8}$He yields are extrapolated from KamLAND for their muon rates, average muon energies, and the detector configurations, and the corresponding rates are 117 and 37 per day per 20 kton, respectively.

The $^{9}$Li/$^{8}$He background stems from the $\beta$-n decays of the isotopes, where the half-lives of $^{9}$Li and $^{8}$He are 0.178 s and 0.119 s, and the branching ratios of their $\beta$-n decay mode are 51\% and 16\%, respectively. The total $\beta$-n decay rate is about 1200 $\rm {yr}^{-1}{kt}^{-1}$. We take the prompt energy spectra of $^{9}$Li and $^{8}$He $\beta$-n decays from Ref.~\cite{An:2015jdp}, {which have the $Q$ values of 11.9 MeV and 8.6 MeV, respectively}.
Finally, we note that the $^{9}$Li/$^{8}$He background can be effectively suppressed by muon veto strategies~\cite{JUNO:2021vlw,JUNO:2020hqc}.
In the end, taking into account all the above considerations, the $^{9}$Li/$^{8}$He background is negligible above a prompt energy of 12 MeV.

\subsection{Fast neutron}\label{sec:bkgfn}

Muons passing through the JUNO LS or through the water buffer will be tagged with almost 100\% and 99.8\% efficiency~\cite{An:2015jdp} respectively. {Neutrons associated with tagged muons can be rejected by muon veto with an efficiency of 100\% and a livetime of 93.6\%~\cite{JUNO:2021vlw,JUNO:2020hqc}. Neutrons associated with untagged muons, which include muons only passing through surrounding rocks and corner clipping muons with the track length in water shorter than 0.5 m,}
might enter the LS and produce a prompt signal before being captured on the proton or carbon with a delayed signal. They contribute to the FN background.
\begin{figure}
	\centering
	\includegraphics[scale=0.6]{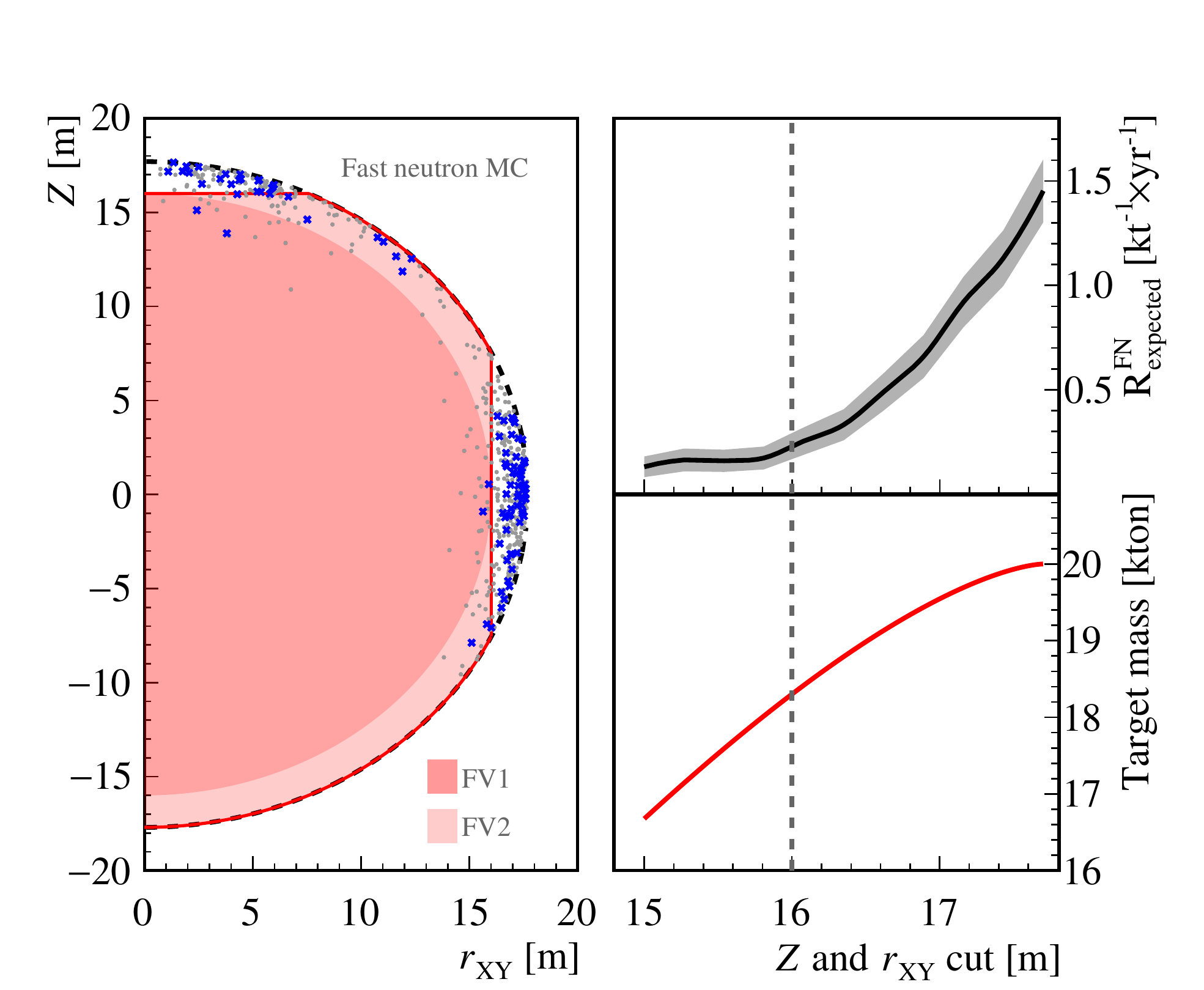}
\caption{ Left panel: Spatial vertex distributions of the simulated FN background, where the dark and light pink colors represent the regions of FV1 and FV2 respectively,
{and the grey and blue points refer to the events in the whole prompt energy range and within [12, 30] MeV respectively}. Right top panel: the event rate of FN background in terms of the $Z$ and {\it r}$_{\rm{XY}}$ cut for the prompt energy of [12, 30] MeV, where the grey band refers to the statistical uncertainty. Right bottom panel: the target mass in terms of the $Z$ and {\it r}$_{\rm{XY}}$ cut.}
	\label{fig:fastn}
\end{figure}

We have performed a muon simulation with the JUNO simulation framework. In order to accelerate the simulating speed, we focus on untagged muons in the surrounding rocks and water pool and neglect the simulation of optical photons. 
Due to the specific geometry of the JUNO detector, most of the FN will be captured at the equator and upper regions of the LS. They can be effectively removed by a fiducial volume cut in terms of the vertical position $Z$ and the horizontal distance to the detector centre ${r}_{\rm{XY}}$. The spatial vertex distributions of the FN events are illustrated in the left panel of Fig.~\ref{fig:fastn}, where the grey and blue points refer to the FN events of the whole prompt energy range and within [12, 30] MeV respectively.
We define two fiducial volume (FV) regions based on the values of {\it Z} and {\it r$_{\rm XY}$}, where the first one (FV1), as shown in the left panel of Fig.~\ref{fig:fastn}, is defined with $R\equiv\sqrt{Z^2+{r}^2_{\rm{XY}}}<16$ m, and the other (FV2) refers to the region with $R>16$ m and $Z<16$ m, $r_{\rm{XY}}<16$ m.
The right top panel of Fig.~\ref{fig:fastn} illustrates the event rate of the FN background as a function of the $Z$ and {\it r}$_{\rm{XY}}$ cut with the prompt energy of [12, 30] MeV, where the grey band is the statistical uncertainty of the simulated data. The corresponding target mass is shown in the right bottom panel of Fig.~\ref{fig:fastn}.
Note that FV2 is designed to enlarge the effective target mass but still avoiding high FN rates. It will be shown in the next section that the efficiencies of the PSD and TC cuts are different for these two regions.
Finally the energy spectrum of the FN background is taken as flat in the selected prompt energy window from 12 to 30 MeV according to the detector simulation outputs.

\subsection{Atmospheric $\nu$ NC background}\label{sec:bkgnc}

The atmospheric neutrino fluxes at JUNO for the neutrino energies from $100~{\rm MeV}$ to $10^{4}~{\rm GeV}$ have been calculated by the Honda group~\cite{Hondahomepage}, where the flux uncertainty is less than $10\%$ in the energy range of $(1-10)~{\rm GeV}$, but gradually increases for both lower and higher energies~\cite{Honda:2015fha,Honda:2019ymh}. For the DSNB analysis, we have performed a systematic study on the CC and NC backgrounds induced by atmospheric neutrino interactions on ${^{12}{\rm C}}$. The CC background is negligible for prompt energies below 100 MeV due to the suppression of neutron production.
The general method of the NC background calculation has been carefully studied in Ref.~\cite{Cheng:2020aaw}. In this work the NC background with the JUNO software framework using full detector simulation has been accomplished to study properties of this important background.

\begin{figure}
	\centering
	\includegraphics[scale=0.7]{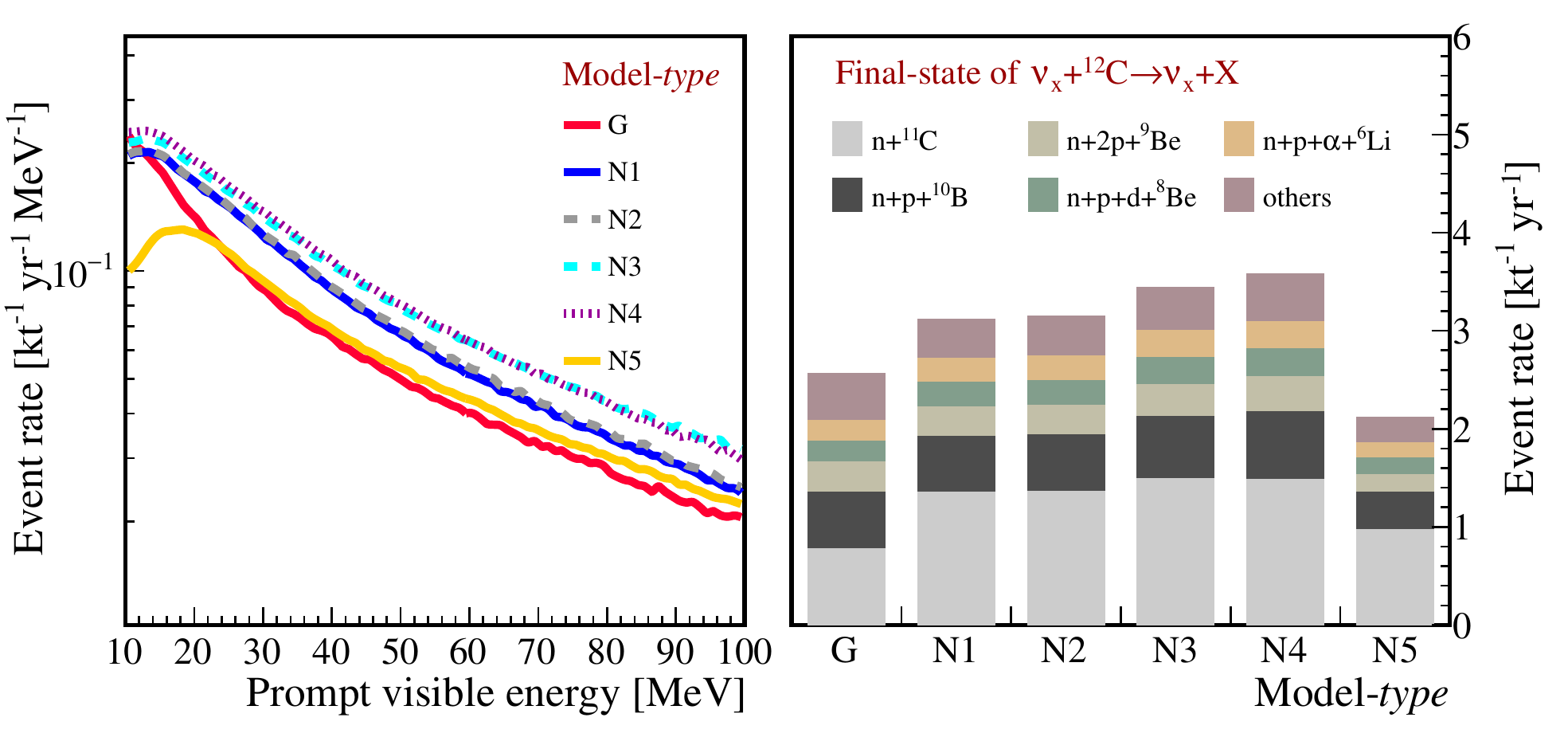}
	\caption{Left panel: event rates of the NC background as a function of the prompt energy using six different neutrino interaction models. Right panel: event rates of the NC background for specific channels with different final-state nuclei in the prompt energy range from 12 to 30 MeV.}
	\label{fig:ncep}
\end{figure}
Two widely-used neutrino generators \texttt{GENIE}~\cite{Andreopoulos:2009rq} and \texttt{NuWro}~\cite{Golan:2012rfa} are used to model the NC interaction between the atmospheric neutrinos and $^{12}$C,
and \texttt{TALYS}~\cite{Koning:2005ezu} is employed to describe deexcitations of the final-state nuclei. Between the two steps, we include a statistical configuration model of $^{12}{\rm C}$ to determine the probability distribution of excited states in the final-state nuclei. Five typical neutrino interaction models have been selected to evaluate the systematic uncertainty of the model prediction, as shown in the left panel of Fig.~\ref{fig:ncep}, where the prompt energy is obtained with the JUNO detector simulation including the full chain of detector response. The first model (G) is from \texttt{GENIE}, and the other five (N$i$ with $i=1,\cdots,5$) are different realization of \texttt{NuWro} with distinct nuclear models and input parameters.
The event rates with different final state nuclei in the prompt energy range from 12 to 30 MeV are illustrated in the right panel of Fig.~\ref{fig:ncep},
where one can notice that the NC background with $^{11}{\rm C}$ is the dominant NC background.
By taking the average of six model calculations as the prediction, and the combination of the flux uncertainty and model variations as the total uncertainty, we arrive at $(3.0 \pm 0.5)~{\rm  kt}^{-1}~{\rm yr}^{-1}$ for the NC background within the prompt energy range from $12$ to $30~{\rm MeV}$.

\begin{figure}
	\centering
	\includegraphics[scale=0.6]{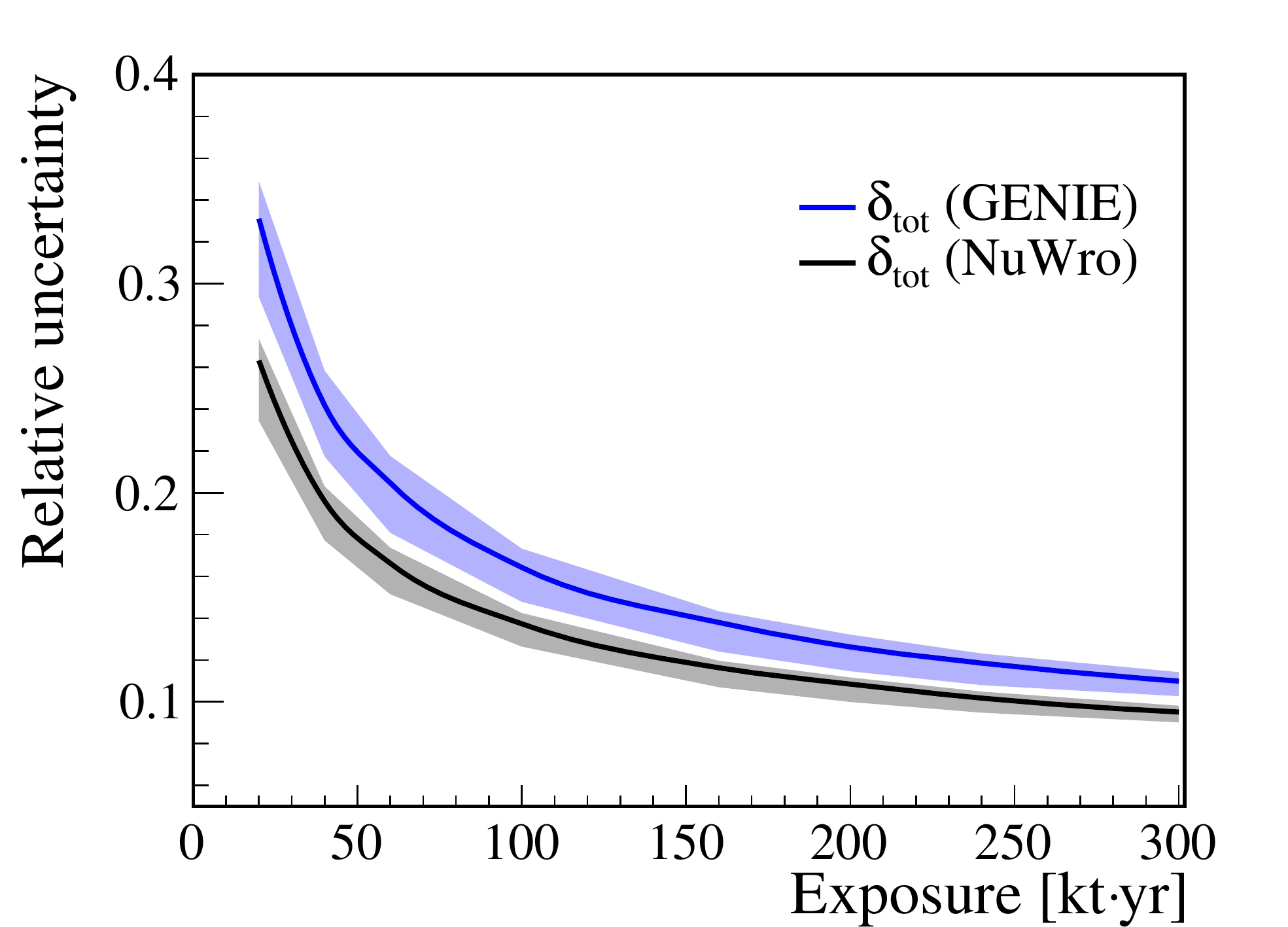}
	\caption{The relative uncertainty of the NC background as a function of the detector exposure by taking account of the envisaged \textit{in situ} measurement at JUNO. The bands are obtained by assuming different levels of natural radioactivity and cosmogenic $^{11}{\rm C}$ in the accidental background.}
	\label{fig:ncunc}
\end{figure}
To test the theoretical prediction and further reduce the uncertainty of the NC background, we can measure the NC background \textit{in situ} with the JUNO detector. In the neutrino NC interactions, some of the final state nuclei, such as $^{11}{\rm C}$ and $^{10}{\rm C}$, may undergo delayed $\beta$ decays, forming a distinct three-fold signature in the detector.
The three-fold signature can be measured with reduced backgrounds and excellent accuracy is shown to be achievable using the JUNO simulation data. 
Then the two-fold NC background is converted from the three-fold signature measurement by using their correlated ratios of model predictions~\cite{Cheng:2020oko}.
Therefore the NC background uncertainty from the \textit{in situ} measurement is obtained with both the statistical and conversion uncertainties,
where the conversion uncertainty is from the model variations of neutrino generators.
The relative uncertainty of the NC background as a function of the detector exposure by using the {\it in situ} measurement is illustrated in Fig.~\ref{fig:ncunc},
where the uncertainty can be decreased from 35\% of one-year data to less than 15\% after around ten years of running. 
The bands are obtained by assuming different levels of natural radioactivity and cosmogenic $^{11}{\rm C}$ in the accidental background.
The difference between \texttt{GENIE} and \texttt{NuWro} is mainly driven by the different branching ratios of the $^{11}{\rm C}$ channel.
In the following calculation, we take the NC background uncertainty as 35\% for the first 3 years of data taking, and an uncertainty of 25\% (15\%) after three (nine) years.

\section{Background suppression}
\label{sec:bkgred}

In this section we discuss the background suppression strategies for the DSNB sensitivity study.
Firstly, we follow the muon veto strategy as studied in Ref.~\cite{JUNO:2021vlw},
where the efficiency of the live time can reach 93.6\%.
Secondly, since different types of particles depositing energies in LS will have distinct photon emission time profiles, the PSD technique will be powerful to distinguish the backgrounds with different profiles of time distributions. Here we present our detailed simulation on the PSD efficiency, and apply for the suppression of the FN and atmospheric $\nu$ NC backgrounds. Finally as mentioned before, the atmospheric $\nu$ NC background associated with the final-state nuclei ${}^{11}\mathrm{C}$ is the most significant background, which undergoes a $\beta^{+}$ decay with a lifetime of 20.39 min and a decay energy of 1.98 MeV. Therefore we make an additional TC cut to effectively reduce this category of the NC background.

\subsection{PSD cut}\label{sec:bkgpsd}

\begin{figure}
	\centering
	\includegraphics[scale=0.60]{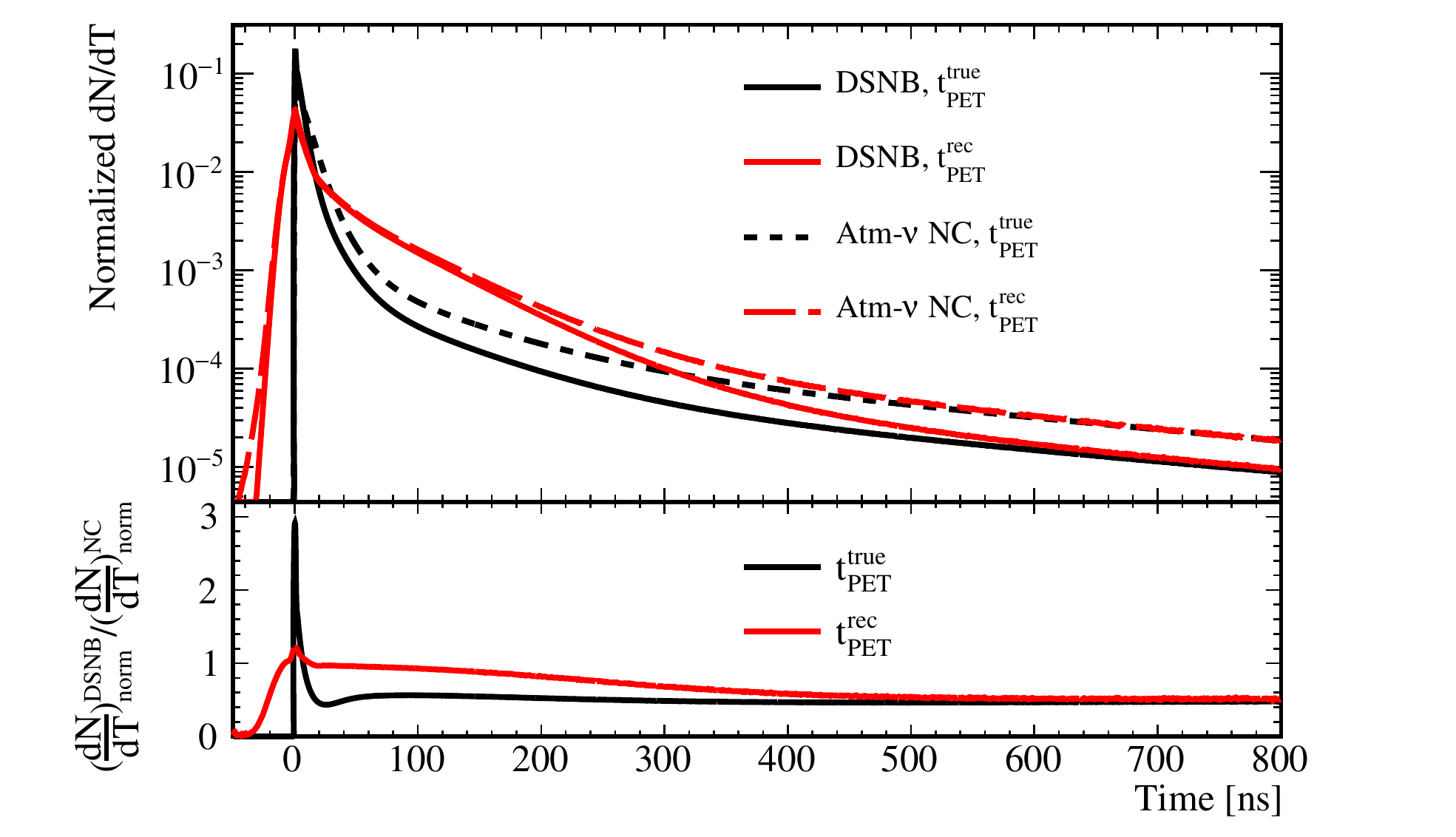}
	\caption{The averaged true (black) and reconstructed (red) profiles of the photon emission time (PET) are illustrated for both the DSNB signal (solid) and NC background (dashed). The upper panel is shown for the normalized time profiles and the lower panel for the relative ratios of the DSNB signal and NC background.}
	\label{fig:timeprofile}
\end{figure}

In organic LS, the fluorescence time profile is characterized by typical decay time constants ranging from several $n$s to several hundred $n$s. The probability of photon emission as a function of time is described by the weighted sums of exponential functions of several components.
The time profiles of different kinds of particles are featured by the distinct time constants and the corresponding weights, which are the foundation of the PSD technique.
During the full detector simulation of the signal or background events, an optical photon starts from the emission time in LS, to the photon propagation before detected by one specific PMT, then it is converted to an electrical signal to be read out and reconstructed.
The simulation is based on the JUNO offline framework, and includes a full chain of the event generator, detector simulation, electronics simulation, waveform reconstruction and event reconstruction. The reference DSNB flux model discussed in Sec.~\ref{sec:dsnb}, the NC background from the \texttt{GENIE} model and the FN background are used to simulate the data. The statistics of the simulation corresponds to around two million events of the signal and background to avoid any bias of the fluctuation.
In Fig.~\ref{fig:timeprofile} the averaged true (black) and reconstructed (red) profiles of the photon emission time (PET) for both the DSNB signal (solid) and NC background (dashed) are illustrated. The upper panel is shown for the normalized time profiles and the lower panel for the relative ratios of the DSNB signal and NC background.
We note that the difference between the true and reconstructed profiles is pretty significant between around 50 and 300 $n$s, which is due to the time-of-flight smearing induced by multiple hits and total reflection.
There are different methods to implement the PSD technique in LS detectors, including the tail-to-total ratio (TTR) method~\cite{Mollenberg:2014pwa}, the multivariate machine learning technique with the Boosted Decision Tree (BDT) option~\cite{Hocker:2007ht}, and the advanced Neural Network (NN) method~\cite{Pedregosa:2012toh}.
In the following, we would like to summarize the general properties of our simulation results on the PSD performance.

Firstly, due to the detector non-uniformity, the performance with the position dependent method is much better than the simple calculation applied to the whole detector. Secondly, since the prompt signal of the NC background contains not only the kinetic energies of nuclei, but also the deposited energies of possible deexcited $\gamma$'s, the BDT method utilizing both the tail and peak signatures surpasses the TTR method that employs the tail information.
The DSNB signal efficiency in the BDT method can reach the level of around 80\% while keeping the residual NC background (denoted as the background inefficiency) as low as 1\%, which will be our baseline option for the sensitivity study. Finally the Scikit-learn toolkit~\cite{Pedregosa:2012toh} is used as an independent NN analysis. By using the same simulation as the BDT method, we show that the NN method can achieve consistent performance for the background suppression, demonstrating the reliability of the PSD efficiencies.

\begin{figure}
	\centering
	\includegraphics[scale=0.40]{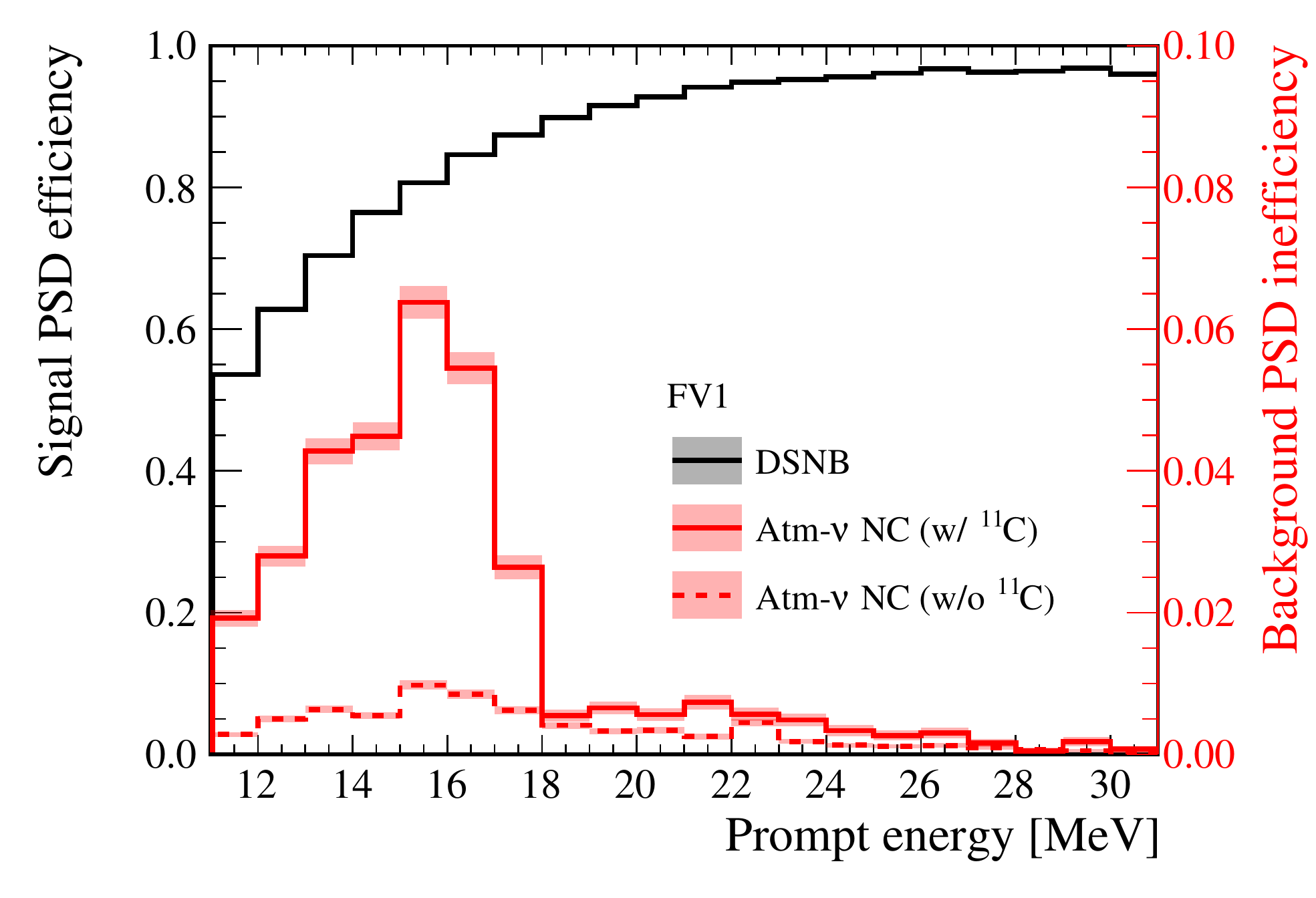}
\hspace{-0.2cm}
	\includegraphics[scale=0.40]{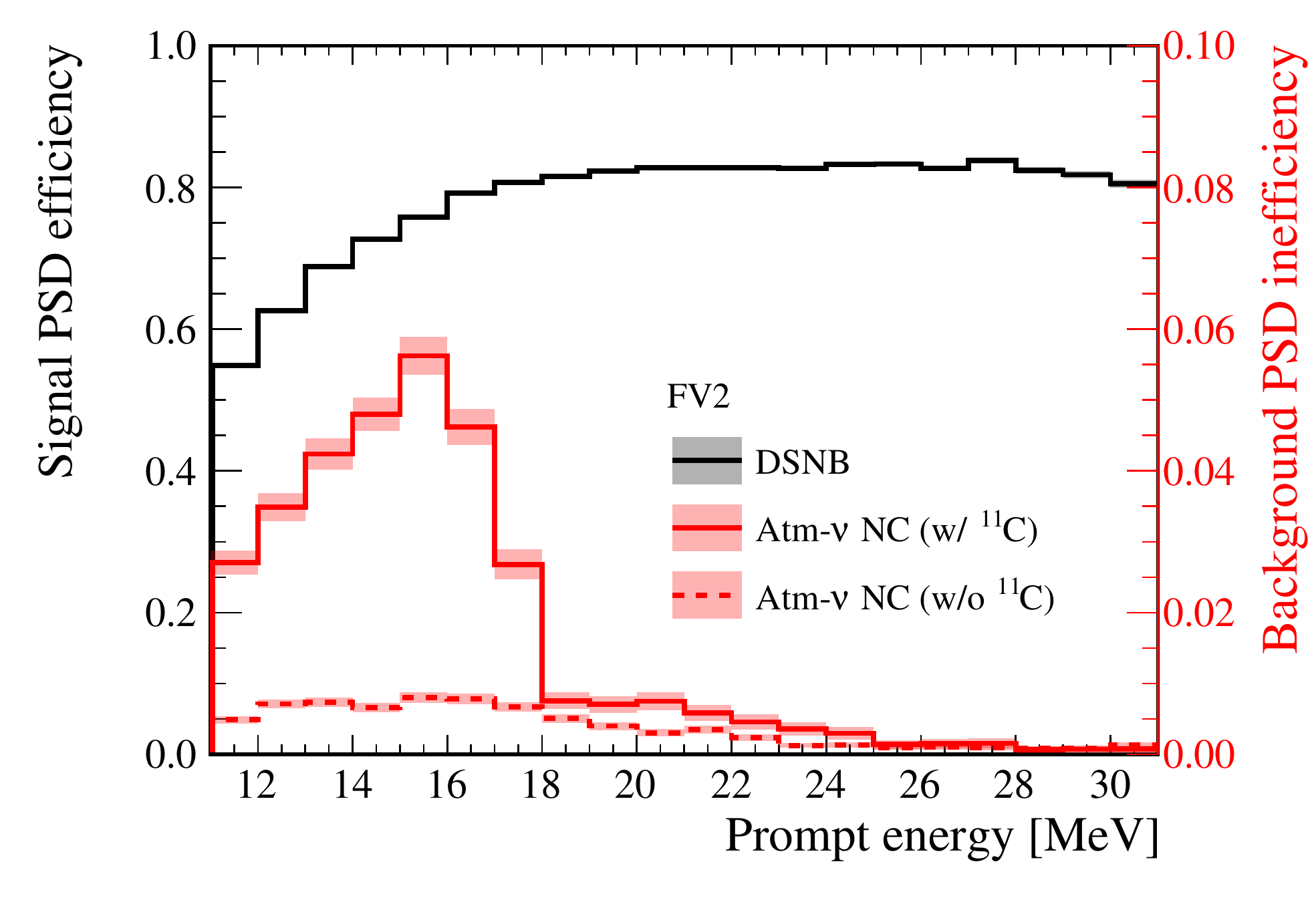}
\vspace{-0.2cm}
	\caption{PSD efficiencies as functions of the prompt energy with the BDT method. The left and right panels are shown for the signal and background efficiencies in the regions of FV1 and FV2. The black solid lines are for the signal efficiency after the PSD cut, and the red lines are for the background inefficiencies of the atmospheric $\nu$ NC backgrounds with (solid) and without (dashed) ${}^{11}\mathrm{C}$.
The shadowed bands are shown for the statistical uncertainty of simulated data samples.}
	\label{fig:psd}
\end{figure}

In Fig.~\ref{fig:psd}, we illustrate the PSD efficiencies as the functions of the prompt energy by requiring the average background inefficiency as 1\% with the BDT method. The left and right panels are for the signal efficiencies and background inefficiencies in the regions of FV1 and FV2 respectively. The black solid lines are for the signal efficiency after the PSD cut, and the red lines are for the background inefficiencies of the atmospheric $\nu$ NC backgrounds with (solid) and without (dashed) ${}^{11}\mathrm{C}$. The shadowed bands are shown for the statistical uncertainty of simulated data samples.
{Note that the choice of the 1\% average background inefficiency has been optimized with higher signal-to-background ratio and better DSNB sensitivity.}

From the figure, several comments can be provided as follows. Firstly, the PSD performance is detector position dependent, the average efficiencies for the DSNB signal are 84\% and 77\% in FV1 and FV2 respectively, where the energy dependence of the signal efficiencies and background inefficiencies is shown in the red and black curves of Fig.~\ref{fig:psd}.
Because of the detector non-uniformity, the total reflection in FV2 would affect the photon time profile and reduce the PSD performance.
Secondly, the PSD efficiencies are particle-type and energy dependent. In Fig.~\ref{fig:psd} we observe that the inefficiencies for the NC background with $^{11}{\rm C}$ are higher than those NC background without $^{11}{\rm C}$, in particular for the events with the prompt energy smaller than 18 MeV, where a sharp increase emerges for both FV1 and FV2 regions.
The NC background with $^{11}{\rm C}$ is pure neutrons with high energies, and the corresponding prompt energy includes both the elastically recoiled protons and other inelastic products from the neutron interactions with $^{12}{\rm C}$. Below a threshold energy at around 18 MeV, the inelastic products are dominant by deexcited $\gamma$'s, and above the threshold the processes with heavy final-state particles become more effective, such as the proton, $\alpha$, $d$, which are relatively easier to recognized in the LS time profile.
For the NC background without $^{11}{\rm C}$, one can also look into the component of the prompt signal, which includes more heavy final-state particles than the NC background with $^{11}{\rm C}$, and thus results in better PSD background rejection power.

Finally we evaluate the associated systematic uncertainty of the PSD cut. Several event samples in future JUNO measurements could be used to directly measure the PSD efficiencies and/or indirectly as inputs of detector simulation tuning.
The first candidate sample is the spallation neutrons, which have similar prompt energies as the DSNB observation window.
The spallation neutrons with muons crossing the outer veto region but without track in the CD can be selected to form a control sample for the NC background.
A detailed study corresponding to around 180 days of muon simulation data has been performed. The event rate of this control sample is around 2 per day in the DSNB search region from 12 to 30 MeV. Considering the average PSD inefficiency of 1\% for the NC background, the statistical uncertainty of this selected sample is at the level of 30\%, 20\%, and 10\% for 1 year, 3 years, and 9 years of data taking, respectively. Note that other control samples including neutron calibration source of the low energy region~\cite{JUNO:2020xtj}, samples of the muon capture and Michael electrons can also be used to control the systematic uncertainty of the PSD cut.

\subsection{TC cut}\label{sec:bkgtc}

The signature of the NC background with ${}^{11}\mathrm{C}$ are three-fold, which typically consists of a prompt signal of the fast neutron recoil, a delayed signal of neutron capture on hydrogen, and an additional signal from beta decay of the unstable ${}^{11}\mathrm{C}$. To optimize the efficiency for the TC cut, we use the same simulation data as in the PSD study and we also consider the accidental coincidence of the muon-induced ${}^{11}\mathrm{C}$ or natural radioactivity with a preceding IBD-like signal. By varying the time and distance between the third delayed signal and the first prompt one, we have obtained an optimal choice for the best sensitivity of the DSNB search, which corresponds to a TC inefficiency of 25.5\% for the NC background with ${}^{11}\mathrm{C}$ and an efficiency of 93.6\% for all the other components. Notice that the optimal TC cut is stable for different detector exposures and the TC cut can only be applied in FV1 because of the rather high background level in FV2.

To summarize this section, in the following DSNB sensitivity study, we use the energy dependent PSD performance in FV1 and FV2 to suppress the NC background.
By splitting the NC background into two categories with and without ${}^{11}\mathrm{C}$, one can also consider an additional TC cut to further suppress the NC background with ${}^{11}\mathrm{C}$ in FV1. It is shown that the PSD performance of the FN background is the same as that of the NC background with ${}^{11}\mathrm{C}$. Finally, we remark that the model of LS scintillation time profiles in this work is particle-type dependent, and thus measurements with low energy events have been used for the DSNB energy range. However, the time profile model and resulting PSD performance may also be energy dependent. In this respect, the efficiencies of the DSNB signal and inefficiencies of the NC background might be revised. We defer this study to a future separated work.

\section{Sensitivity}\label{sec:sen}

	\begin{table}[htbp]
	\centering	
	\begin{tabular}{l|c|cc|cc|cc}
		\hline	
		\hline	
		{\bf{Signal }} & \bf{Rate}[147 $\text{kt} \times \text{yr}$]  & \multicolumn{2}{c|}{muon veto}  &
		\multicolumn{2}{c|}{PSD}  & \multicolumn{2}{c}{TC cut}  \\
		\hline	
		12 MeV & 16.2  &
		\multirow{4}{*}{93.6\%} & 15.2
		& \multirow{4}{*}{ }  & 12.9
		& \multirow{4}{*}{93.6\%} & 12.1 \\
		15 MeV & 20.8  &  & 19.4  &   & 16.7  &  & 15.6 \\
		18 MeV & 25.2  &  & 23.6  &   & 20.4  &  & 19.1\\
		21 MeV & 29.0  &  & 27.2  &   & 23.7  &  & 22.1 \\
		\hline	
		{\bf{Backgrounds }} &       &       &       &       &       &       &  \\
		\hline	
		Fast neutron    & 12.5  &  & 11.7   &    & 0.2   &  & 0.2 \\
		Atm-$\nu$ CC    & 2.0   &  & 1.9   &   & 1.6   &  & 1.5 \\
		Atm-$\nu$ NC without ${}^{11}\mathrm{C}$
		& 258.2   &  & 241.7 &    & 0.9   &  & 0.9 \\
		Atm-$\nu$ NC with ${}^{11}\mathrm{C}$
		& 186.7 &  & 174.8 &    & 3.6   & 25.5\% & 0.9 \\
		\hline
		{\bf{Total backgrounds }} & 459.4 &       & 430.0 &       & 6.3   &       & 3.5 \\
		\hline	
		\hline	
	\end{tabular}%
	\caption{\label{ts2} Event rates of the DSNB signal and corresponding backgrounds in FV1 with the prompt energy in [12, 30] MeV. For the DSNB signal, we have assumed the black hole fraction of 0.27, the SN rate at $z=0$ of $ 1.0 \times 10^{-4} \, \text{yr}^{-1}\text{Mpc}^{-3}$, and four different SN average energies of 12, 15, 18 and 21 MeV.}
\end{table}%
\begin{table}[htbp]
	\centering	
	\begin{tabular}{l|c|cc|cc}
		\hline	
		\hline	
		{\bf{Signal }} & \bf{Rate}[36 $\text{kt} \times \text{yr}$]  & \multicolumn{2}{c|}{muon veto}  &
		\multicolumn{2}{c}{PSD}    \\
		\hline	
		12 MeV & 3.9  &
		\multirow{4}{*}{93.6\%} & 3.6
		& \multirow{4}{*}{ }  & 2.8     \\
		15 MeV &5.0   & & 4.6      & & 3.6  \\
		18 MeV &6.0   & & 5.6      & & 4.4  \\
		21 MeV &6.9   & & 6.5      & & 5.1  \\
		
		\hline	
		{\bf{Backgrounds }} &       &       &       &       &         \\
		\hline	
		Fast neutron  & 31.2  &  & 29.2  & & 0.5     \\
		Atm-$\nu$ CC & 0.5   & & 0.4   &  & 0.4     \\
		Atm-$\nu$ NC without ${}^{11}\mathrm{C}$ &
		62.5  &       & 58.5  &       & 0.2     \\
		Atm-$\nu$ NC with ${}^{11}\mathrm{C}$ &
		42.3  &       & 39.6  &       & 0.8     \\
		\hline
		{\bf{Total backgrounds }} & 136.5 &       & 127.8 &       & 1.9   \\
		\hline	
		\hline	
	\end{tabular}%
	\caption{\label{ts3} The same as Tab.~\ref{ts2} but for the region of FV2. Note that the PSD efficiencies are different and the TC cut is not applied for FV2.}
\end{table}%
\begin{figure}
	\centering
		\includegraphics[width=0.9\textwidth]{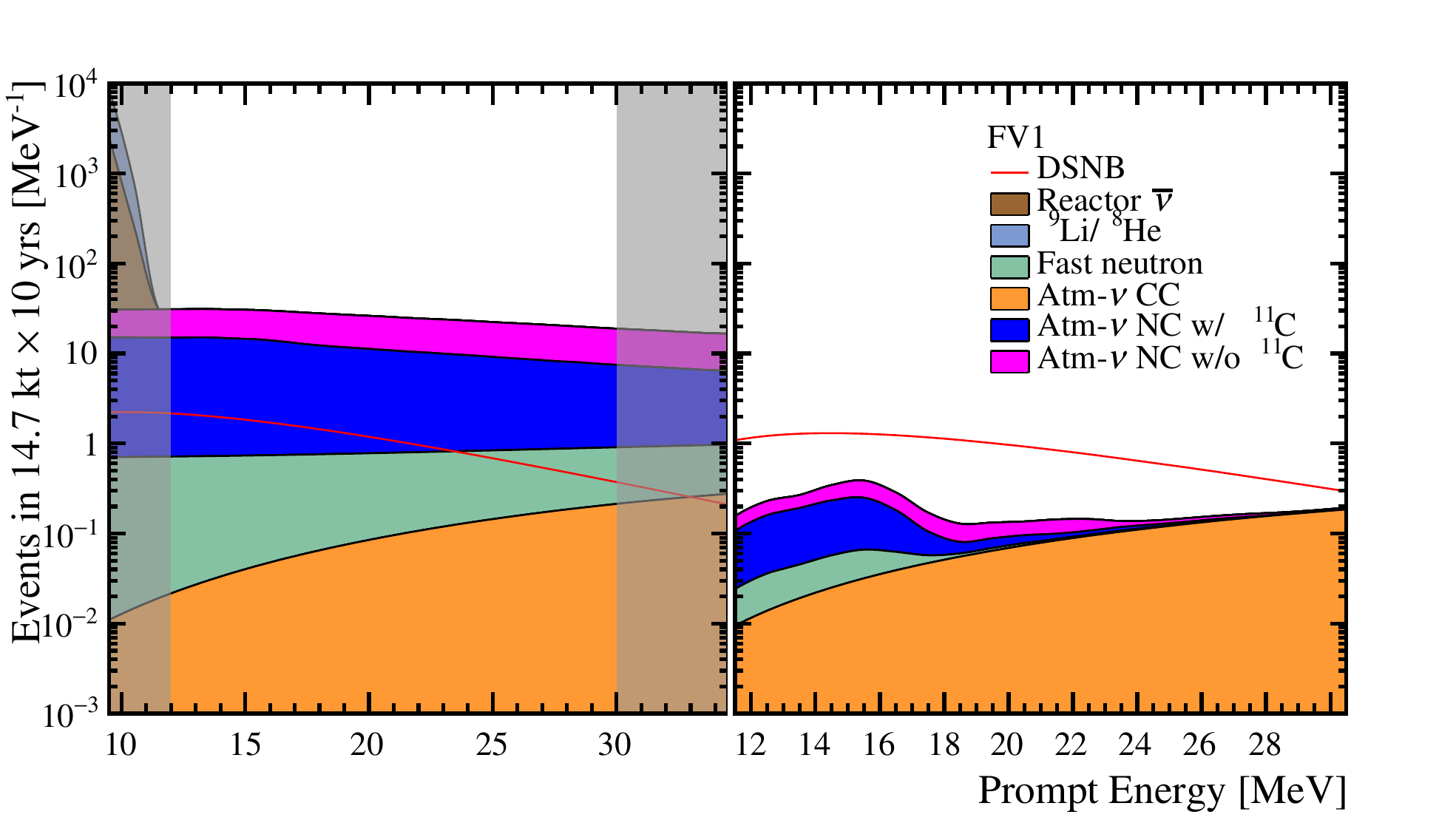}
	\hspace{-1in}
		\includegraphics[width=0.9\textwidth]{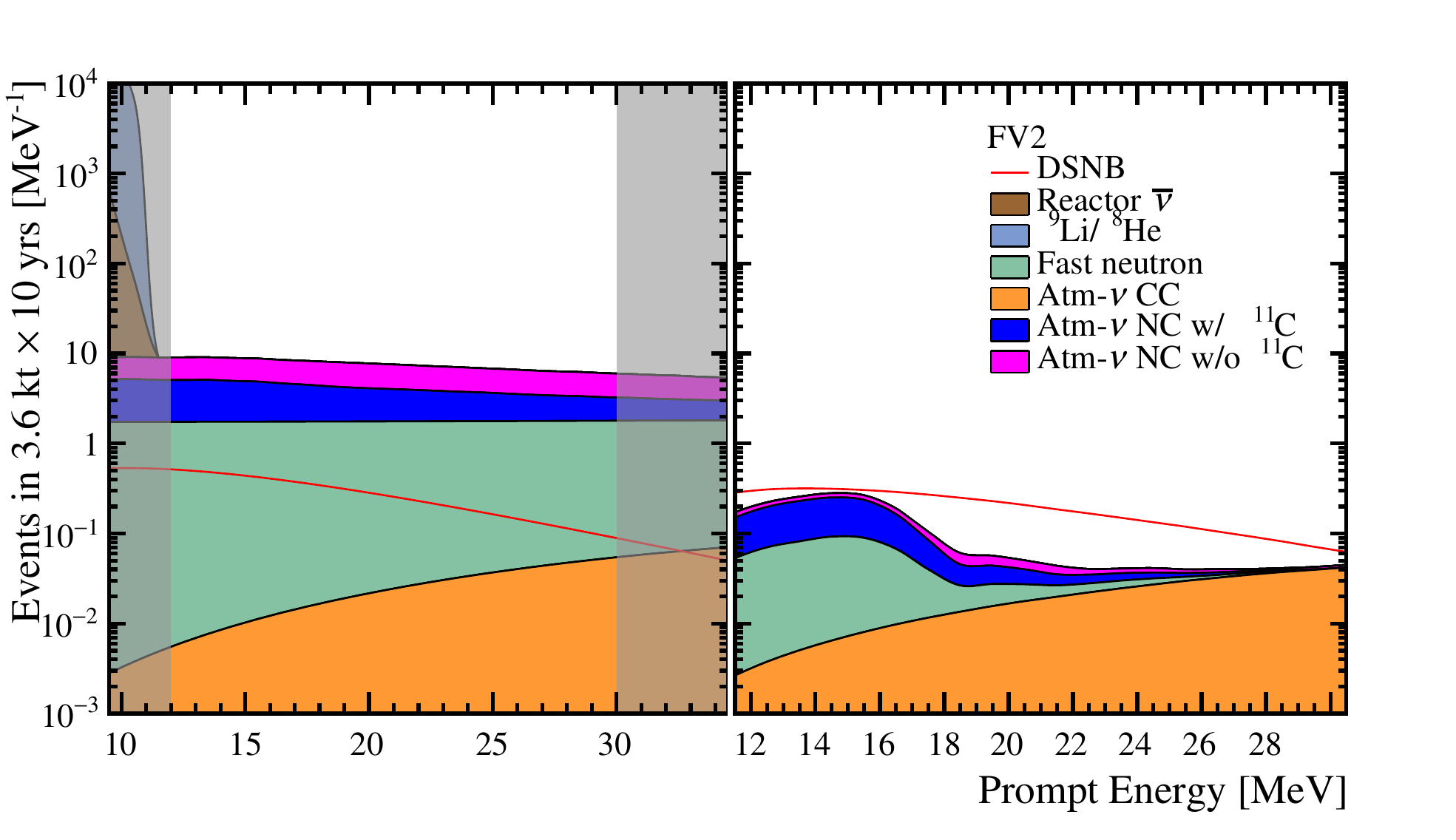}
	\setlength{\abovecaptionskip}{-0.0cm}
	\caption{The prompt energy spectra of the reference DSNB signal with $R_{\mathrm{SN}}(0)=1.0 \times 10^{-4} \, \text{yr}^{-1}\text{Mpc}^{-3}$, $\langle E_{\nu}\rangle=15\,\mathrm{MeV}$, and $f_\text{BH} = 0.27$ versus all the backgrounds before (left) and after (right) the background reduction techniques. The upper and lower panels are shown for the regions of FV1 and FV2 respectively.}
	\label{fig:signalspc}
\end{figure}

In this section we discuss the DSNB sensitivity at JUNO.
To begin with, we provide a summary of the DSNB signal and background evaluations.
The event rates of the signal and background for 10 years of data taking are given in Tab.~\ref{ts2} and Tab.~\ref{ts3} for the fiducial regions of FV1 and FV2 respectively, where we have assumed the black hole fraction of 0.27 and the present SN rate of $ 1.0 \times 10^{-4} \, \text{yr}^{-1}\text{Mpc}^{-3}$. The signal rates with different average energies of SN neutrinos are provided for the prompt energy within [12, 30] MeV, {where both the lower and higher boundaries of the prompt energy range have been optimized with the DSNB discovery potential for the whole DSNB parameter space.}
The signal and background rates using the background reduction techniques of muon veto, the energy dependent PSD cut and the TC cut (only in FV1) are also shown in the tables.
Meanwhile, the prompt energy spectra of the reference DSNB signal with $R_{\mathrm{SN}}(0)=1.0 \times 10^{-4} \, \text{yr}^{-1}\text{Mpc}^{-3}$, $\langle E_{\nu}\rangle=15\,\mathrm{MeV}$, and $f_\text{BH} = 0.27$ and all the backgrounds before (left) and after (right) the background reduction techniques are illustrated in Fig.~\ref{fig:signalspc}. The upper and lower panels are shown for the regions of FV1 and FV2 respectively.
We can notice that after the background suppression, the DSNB signal becomes visible in the prompt energy window between 12 to 30 MeV.

In order to calculate the DSNB sensitivity, we employ the {Poisson}-type log-likelihood ratio (denoted as $\chi^{2}$) as our test statistics:
\begin{equation}\label{dsnbchi2}
\chi^{2}(\langle E_{\nu} \rangle, f_{\rm BH}, R_{\rm SN}(0)) =
\sum_{i}^{}-2\log\left[P\left(n_{i}, \Phi s_{i} + \sum_{j}f_{j}b_{j,i}\right)\right]
+\sum_{j}\dfrac{(f_{j}-1)^{2}}{\sigma_{j}^{2}}
\end{equation}
where, $P$ is the {Poisson} probability to obtain $n_{i}$ events in the $i$-th bin based on the signal prediction $s_{i}$ and background $b_{j,i}$ with $j$ being the background index. $\Phi$ and $f_{j}$ are the spectral normalization of the signal and backgrounds, respectively, where $\sigma_{j}$ are the systematic uncertainties, which have been specified in the previous section. In this work, the Asimov data set is used to derive the median sensitivity.
The DSNB discovery sensitivity ($\sigma$) is defined as square root of the difference between minimal values of $\chi^{2}$ with ($\Phi = 1 $) and without ($\Phi = 0 $) the DSNB signal after marginalization of other parameters:
\begin{equation}\label{dsnbsensitivity}
\sigma=\sqrt{\Delta \chi^{2}_{\min}} =\sqrt{
\lvert\chi^{2}_{\min}(\Phi=0)
-\chi^{2}_{\min}(\Phi=1) \rvert}
\end{equation}
The discovery sensitivity is a function of the DSNB physical parameters, where we have taken as the SN rate, the SN average energy, and the black hole fraction.

\begin{figure}
	\centering
	\includegraphics[width=0.9\textwidth]{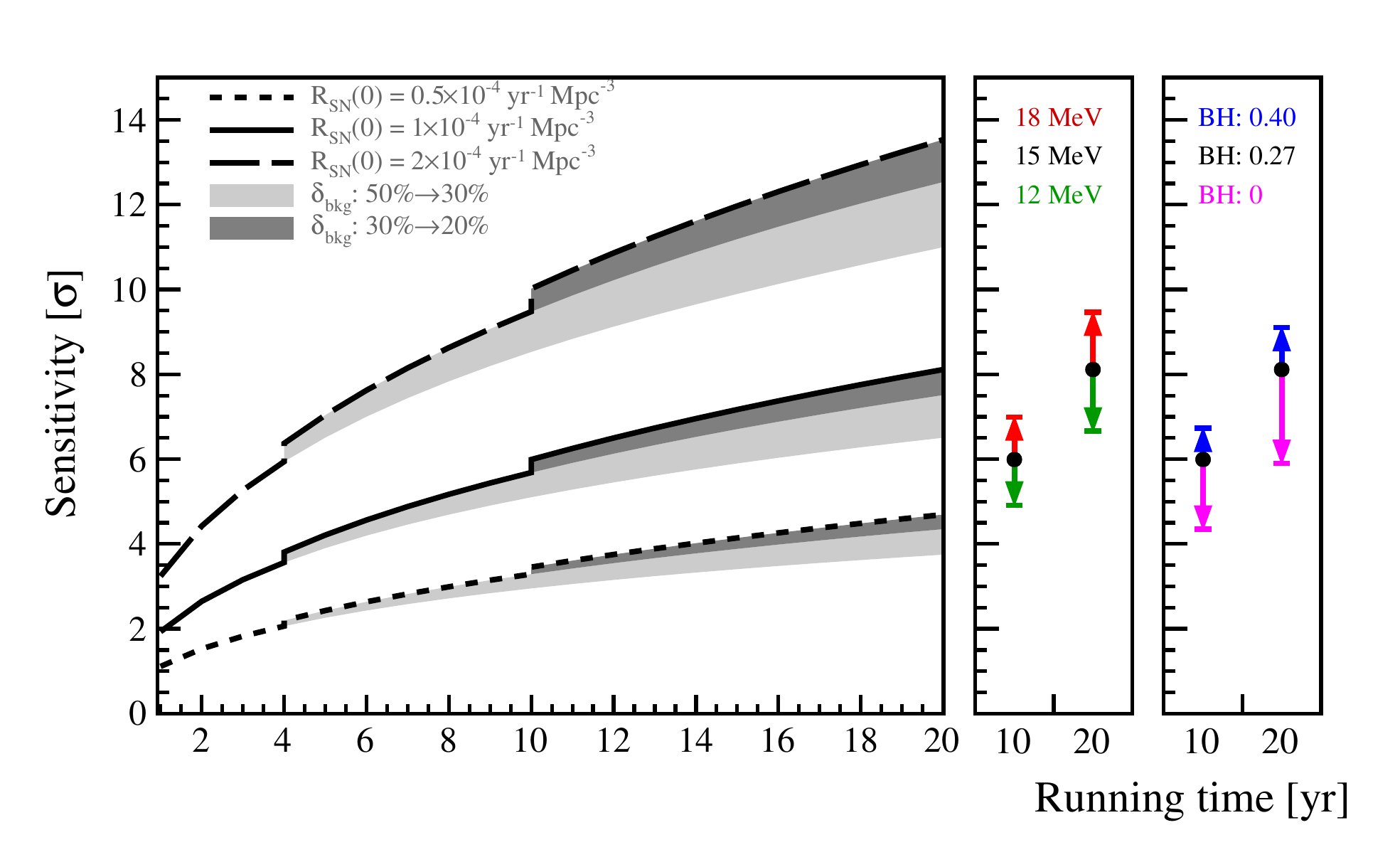}
	\caption{DSNB discovery potential ($\sigma$) at JUNO as a function of the running time.
The reference DSNB signal model is represented with black solid line in the left panel and black circle points in the middle and right panels respectively.
In the left panel, the model variations with represented SN rates from 0.5 to $2.0\times10^{-4} \rm yr^{-1} Mpc^{-3}$ are adopted by using short dashed and long dashed lines respectively. The dark grey and grey regions are illustrated for different choices of the systematic uncertainty of the NC background. In the middle and right panels, the model variations for the SN average energy from 12 to 18 MeV (middle) and the black hole fraction from 0 to 0.40 (right) are illustrated for 10 and 20 years of data taking.}
	\label{fig:sentime}
\end{figure}
In Fig.~\ref{fig:sentime} we illustrate the DSNB discovery potential at JUNO as a function of the running time.
The reference DSNB signal model is taken as $R_{\mathrm{SN}}(0)=1.0 \times 10^{-4} \, \text{yr}^{-1}\text{Mpc}^{-3}$, $\langle E_{\nu}\rangle=15\,\mathrm{MeV}$, and $f_\text{BH} = 0.27$, which is represented with black solid line in the left panel and black circle points in the middle and right panels respectively.
In the left panel, the model variations with represented SN rates from 0.5 to $2.0\times10^{-4} \rm yr^{-1} Mpc^{-3}$ are adopted by using short dashed and long dashed lines respectively. The dark grey and grey regions are illustrated for different choices of the systematic uncertainty of the NC background, which, by the quadratic combination of the uncertainties from the {\it in situ} measurement and the PSD cut, is taken as 50\%, 30\% and 20\%, for 1-3 years, 4-9 years and 10-20 years of data taking, respectively.
In the middle and right panels, the model variations for the SN average energy from 12 to 18 MeV (middle) and the black hole fraction from 0 to 0.40 (right) are illustrated for 10 and 20 years of data taking.
From the figure we can conclude that, for the reference DSNB signal model, JUNO can achieve the sensitivity of $3\sigma$ for around 3 years of data taking and better than $5\sigma$ after ten years.
The discovery potential will increase for higher SN rates, larger SN average energies, and greater black hole fraction, where even for the most pessimistic DSNB model, the sensitivity will arrive at the level of $3\sigma$ for 10 years of data taking.

\begin{figure}
	\centering
	\includegraphics[width=0.8\textwidth]{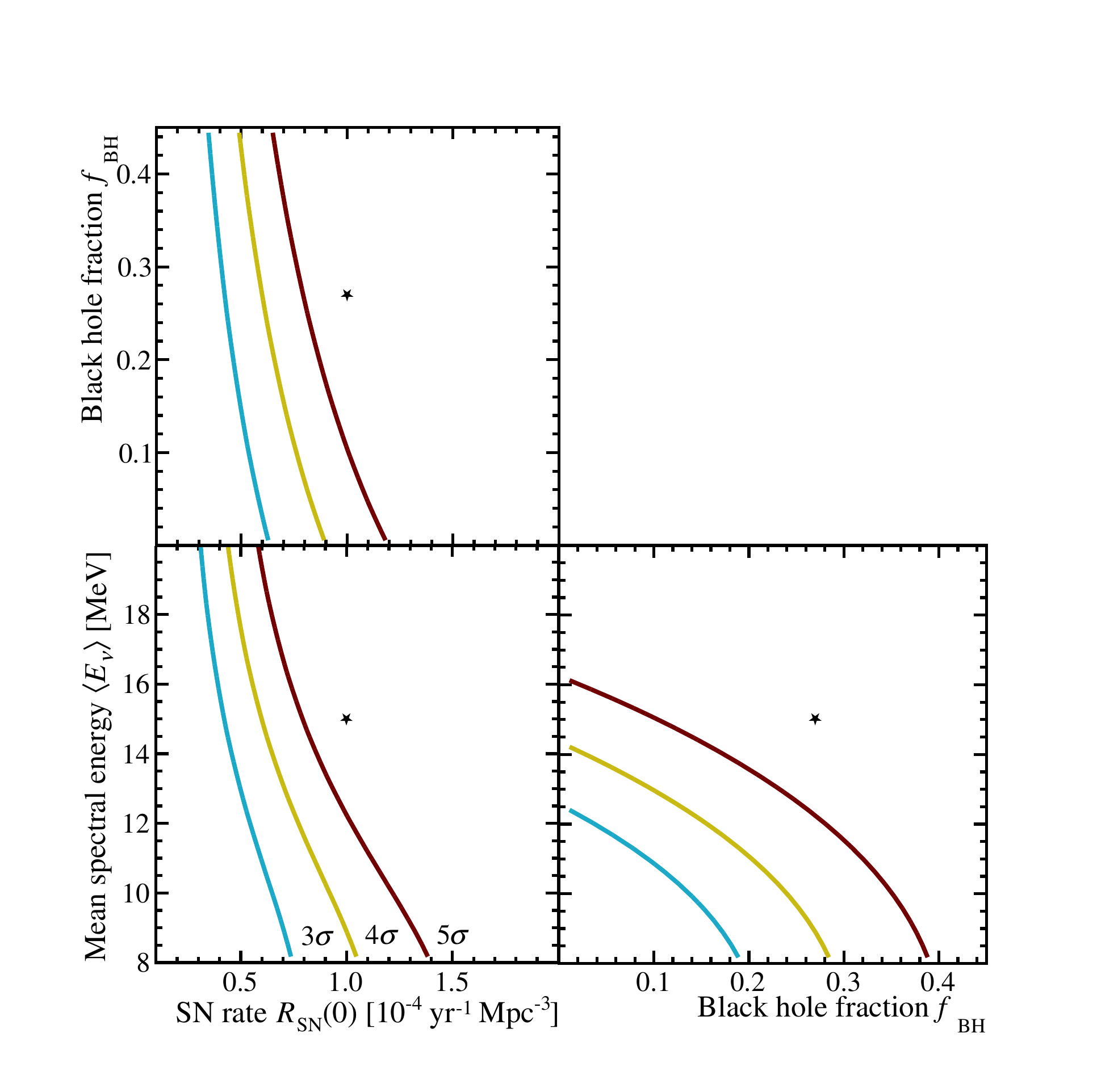}
	\vspace{-0.2in}
	\caption{DSNB discovery potential ($\sigma$) at JUNO as a function of DSNB model parameters for ten years of data taking. The bottom left plot shows the plane of ($R_\text{SN}(0), \langle E_\nu \rangle$, with $f_\text{BH} = 0.27$, the bottom right plot shows the plane of ($f_\text{BH},\langle E_\nu \rangle$) with $R_\text{SN}(0) =1.0\times10^{-4} \rm yr^{-1} Mpc^{-3}$ and the top left plot shows the plane of ($R_\text{SN}(0), f_\text{BH}$) plane with $\langle E_\nu \rangle = 15$ MeV. The blue, yellow and red curves stand for 3$\sigma$, 4$\sigma$ and 5$\sigma$ confidence levels respectively.
The black stars of better than 5$\sigma$ discovery potential show the locations of the reference DSNB model.
	}
	\label{fig:sen2d}
\end{figure}
To further illustrate the model dependence of the DSNB sensitivity, we illustrate in Fig.~\ref{fig:sen2d} the DSNB discovery potential as a function of model parameters for ten years of data taking, where the bottom left plot shows the plane of ($R_\text{SN}(0), \langle E_\nu \rangle$) with $f_\text{BH} = 0.27$, the bottom right plot shows the plane of ($f_\text{BH},\langle E_\nu \rangle$) with $R_\text{SN}(0) =1.0\times10^{-4} \rm yr^{-1} Mpc^{-3}$ and the top left plot shows the plane of ($R_\text{SN}(0), f_\text{BH}$) plane with $\langle E_\nu \rangle = 15$ MeV. These two-dimensional plots with two degrees of freedom are obtained after the marginalization of all the nuisance parameters. The blue, yellow and red curves stand for the 3$\sigma$, 4$\sigma$ and 5$\sigma$ confidence levels of the discovery potential respectively.
The black stars of better than 5$\sigma$ discovery potential show the locations of the reference DSNB signal model.
Comparing to the results of JUNO (2015) in Ref.~\cite{An:2015jdp}, we can conclude that with the latest DSNB signal prediction, more realistic background evaluation and PSD efficiency optimization, and additional TC cut, even greater discovery potential can be obtained for the DSNB observation at JUNO.

If there is no positive DSNB detection, JUNO can also significantly improve the current best limits on the DSNB fluxes.
Assuming the observation equals to the expected background, there are two different and complimentary ways to report the exclusion limits.

\begin{figure}
	\centering
	\includegraphics[width=0.65\textwidth]{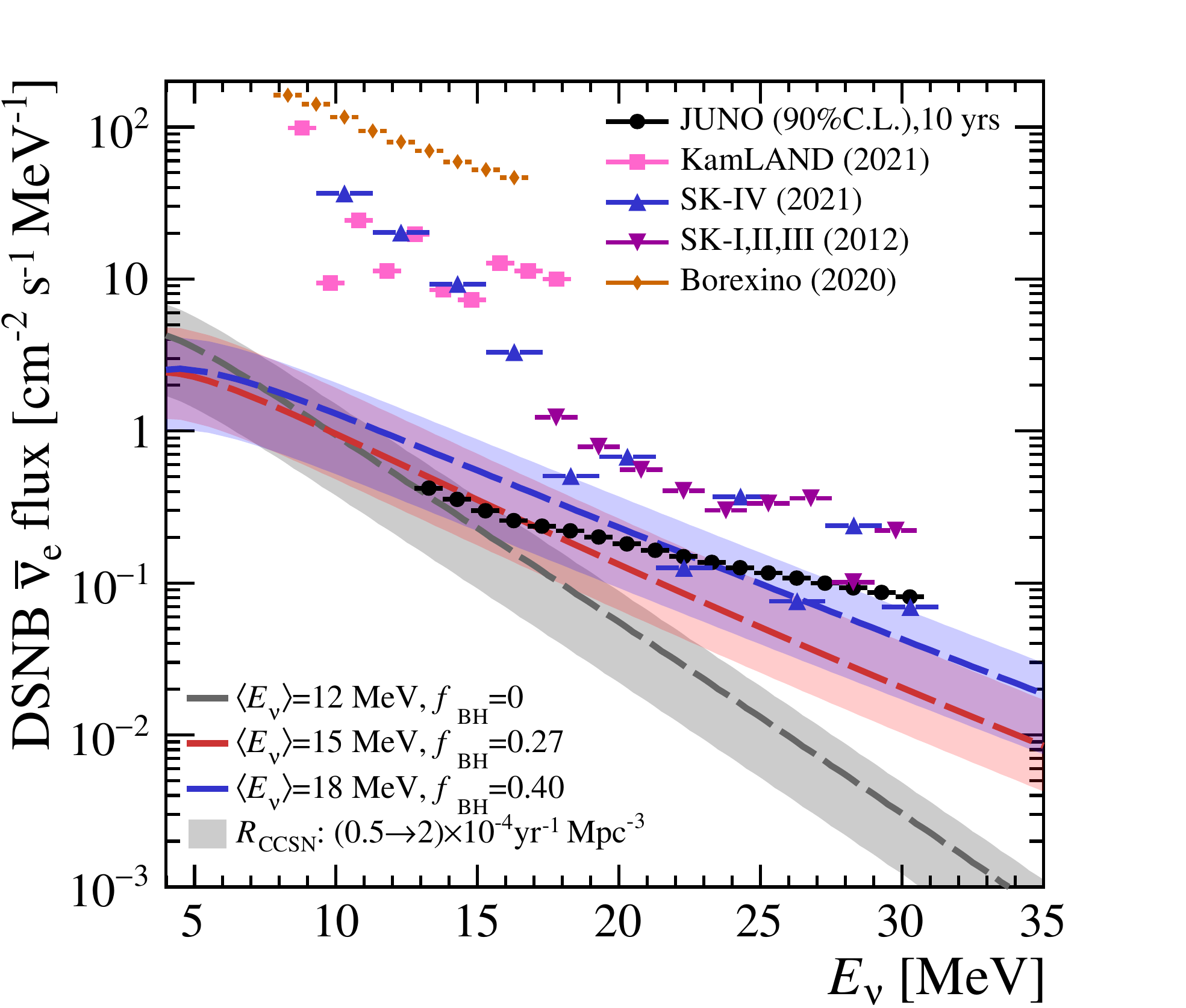}
	\caption{90\% confidence level upper limits on the DSNB fluxes for 18 equal neutrino energy bins from 12 to 30 MeV. The grey, red and blue bands with dashed lines are shown for the DSNB flux predictions with ($\langle E_\nu \rangle=12$ MeV, $f_\text{BH}=0$), ($\langle E_\nu \rangle=15$ MeV, $f_\text{BH}=0.27$) and ($\langle E_\nu \rangle=18$ MeV, $f_\text{BH}=0.40$) respectively. The width of these three bands are taken with $R_\text{SN}(0)$ ranging from 0.5 to $2.0\times10^{-4} \rm yr^{-1} Mpc^{-3}$.
The red and blue triangle points are shown for the DSNB flux limits obtained from SK-I,II,III~\cite{Bays:2011si} and SK-IV~\cite{Super-Kamiokande:2021jaq} respectively. The pink square points are taken from the KamLAND detection limits~\cite{KamLAND:2021gvi}. The orange diamond points are shown for the limits from Borexino~\cite{Borexino:2019wln}.}
	\label{fig:limitFC}
\end{figure}
The first method is to select a small energy window and directly derive the upper limit of the DSNB flux in this window using the rate counting method and the Feldman-Cousins statistics~\cite{Feldman:1997qc}.
In Fig.~\ref{fig:limitFC}, we derive the 90\% confidence level upper limits on the DSNB fluxes for 18 equal energy bins from 12 to 30 MeV. The grey, red and blue bands with dashed lines are shown for the DSNB flux predictions with ($\langle E_\nu \rangle=12$ MeV, $f_\text{BH}=0$), ($\langle E_\nu \rangle=15$ MeV, $f_\text{BH}=0.27$) and ($\langle E_\nu \rangle=18$ MeV, $f_\text{BH}=0.40$) respectively. The width of these three bands are taken with $R_\text{SN}(0)$ ranging from 0.5 to $2.0\times10^{-4} \rm yr^{-1} Mpc^{-3}$.
The red and blue triangle points are shown for the DSNB flux limits obtained from SK-I,II,III~\cite{Bays:2011si} and SK-IV~\cite{Super-Kamiokande:2021jaq} respectively. The pink square points are taken from the KamLAND detection limits~\cite{KamLAND:2021gvi}.
The orange diamond points are shown for the limits from Borexino~\cite{Borexino:2019wln}.
From the figure, we can observe that it is very promising for JUNO to reach the parameter space of the DSNB model in the whole neutrino energy range from 12 to 30 MeV.
In the low energy part, it can improve the KamLAND and Borexino limits by around two orders of magnitude. Compared to the SK limit, the improvement is also significant, from one order of magnitude for low energy bins to around three times near the high energy boundary.
It should be noted that the advantage of this method is totally model-independent and much conservative, where only the background budgets are required in the analysis.

\begin{figure}
	\centering
\includegraphics[width=0.65\textwidth]{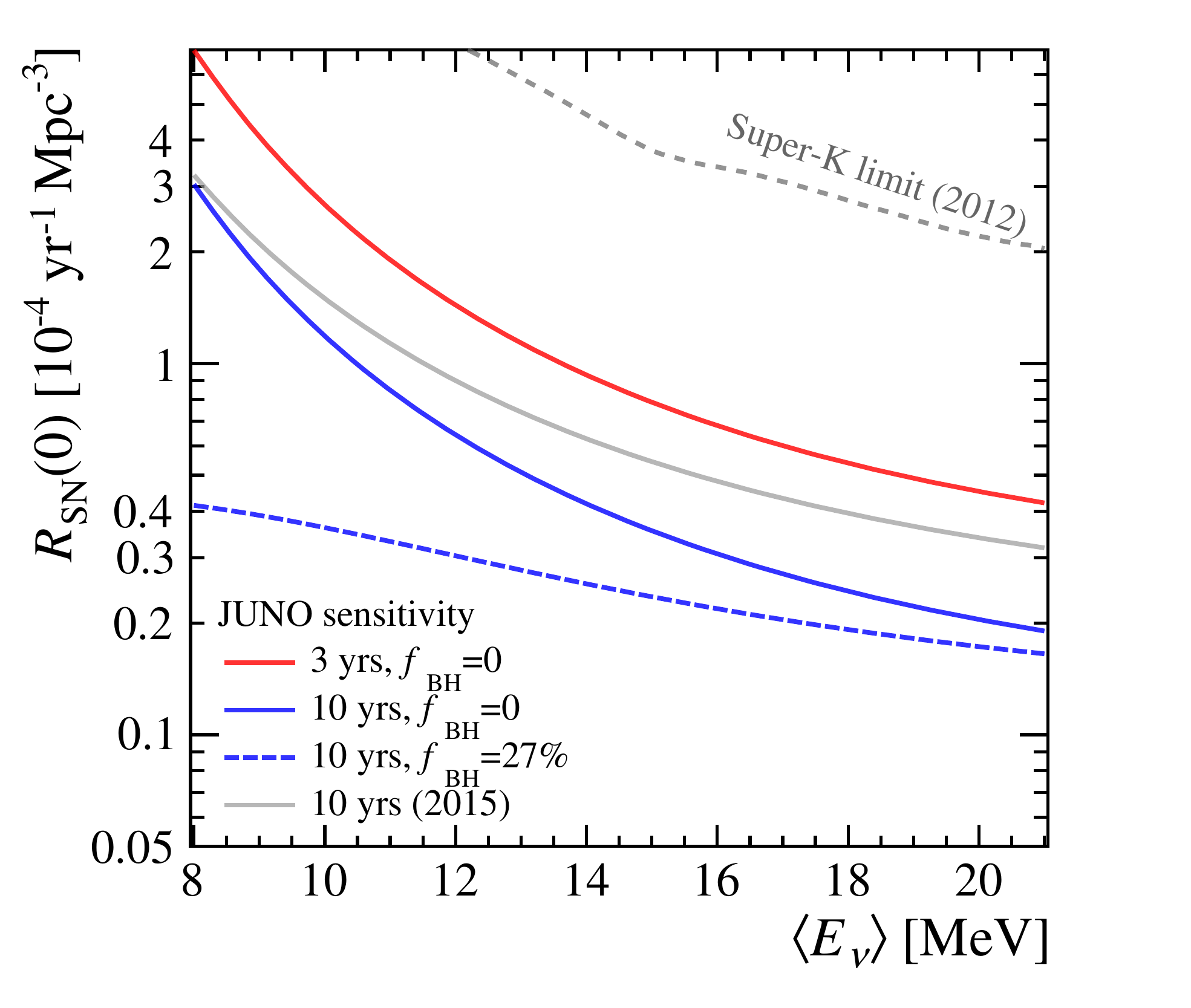}
	\setlength{\abovecaptionskip}{-0.0cm}
	\caption{90\% confidence level upper limits on the DSNB signal in terms of the present SN rate $R_\text{SN}(0)$ as a function of the average energy of SN neutrinos at JUNO. The solid red and blue lines are shown for the limits with $f_\text{BH} = 0$ and running time of 3 years and 10 years, respectively.
The dashed blue line is the limit with $f_\text{BH} = 0.27$ and 10 years of data taking.
{The solid grey line is reproduced for the results of JUNO (2015) in Ref.~\cite{An:2015jdp}} and the dashed grey line is reproduced for the current best limit from SK~\cite{Bays:2011si,Zhang:2013tua}.}
	\label{fig:limit2d}
\end{figure}
Another complementary method of setting the upper limits is to assume a DSNB flux model and use the spectral analysis. To illustrate we simplify the DSNB flux model by fixing the parameter of $f_\text{BH}$, and show in Fig.~\ref{fig:limit2d} the 90\% confidence level upper limits on the DSNB signal in terms of the present SN rate $R_\text{SN}(0)$ as a function of the average energy of SN neutrinos at JUNO. The solid red and blue lines are shown for the limits with $f_\text{BH} = 0$ and running time of 3 years and 10 years, respectively.
The dashed blue line is the limit with $f_\text{BH} = 0.27$ and 10 years of data taking.
{The solid grey line is reproduced for the results of JUNO (2015) in Ref.~\cite{An:2015jdp}} and
the dashed grey line is reproduced for the current best limit from SK~\cite{Bays:2011si,Zhang:2013tua}.
By using 3 years and 10 years of data taking, one can observe that JUNO can significantly improve the current best limit~\cite{Bays:2011si,Zhang:2013tua} by a factor of five and ten respectively.
Compared to the results of JUNO (2015) in Ref.~\cite{An:2015jdp}, the total target mass of FV1 and FV2 is comparable to that of 17 kt, but the PSD efficiencies are improved from an energy independent value of 50\% to the energy dependent efficiencies of around 80\% in this work. Other updates include the efficiencies of muon veto and the TC cut which are both neglected in Ref.~\cite{An:2015jdp}. Considering all these updates, the current work with $f_\text{BH} = 0$ has improved the exclusion limit by 70\% for large average energies and by 40\% for the average energy at around 12 MeV.
Meanwhile, by comparing the blue solid and dashed lines, we observe that the inclusion of the black hole forming SN with nonzero $f_\text{BH}$ would further improve the exclusion limit, which is even more significant for smaller average energies of SN neutrinos.


\section{Concluding remarks}

Large LS detectors are one of the most powerful tools to detect the long-awaited DSNB signal.
In this work, we have made a comprehensive study on the prospects for detecting the DSNB signal at JUNO using the IBD detection channel on free protons.
We have employed the latest DSNB signal predictions based on sophisticated SN numerical simulation, and investigated in great detail the background evaluation and reduction techniques for the DSNB observation.
We have stressed that the atmospheric $\nu$ induced NC background is the most critical background, and demonstrated the powerful PSD technique and excellent TC cut can effectively suppress the NC background and achieve promising discovery potential of the DSNB signal. For the reference DSNB model, JUNO can reach the significance of 3$\sigma$ for around 3 years of data taking, and better than 5$\sigma$ after 10 years. Even for the pessimistic scenario with non-observation, JUNO would strongly improve the current best limits and exclude a significant region of the model parameter space.

JUNO will finish the detector construction and start the journey to the DSNB detection in 2023. Together with the existing water-Cherenkov detector SK-Gd, it stands for the pioneering efforts to first observe the DSNB signal in the next decade~\cite{Li:2022myd}.
In the far future, in order to achieve the goal of doing neutrino astronomy and cosmology with the DSNB observation~\cite{Moller:2018kpn,deGouvea:2020eqq}, one would rely on high-statistics observation with future large-scale detectors, such as Hyper-Kamiokande~\cite{Abe:2018uyc}, DUNE~\cite{DUNE:2020ypp}, LENA~\cite{Wurm:2011zn} and the water-based LS detector THEIA~\cite{Askins:2019oqj}.


\section*{Acknowledgements}
We are grateful for the ongoing cooperation from the China General Nuclear Power Group.
This work was supported by
the Chinese Academy of Sciences,
the National Key R\&D Program of China,
the CAS Center for Excellence in Particle Physics,
Wuyi University,
and the Tsung-Dao Lee Institute of Shanghai Jiao Tong University in China,
the Institut National de Physique Nucl\'eaire et de Physique de Particules (IN2P3) in France,
the Istituto Nazionale di Fisica Nucleare (INFN) in Italy,
the Italian-Chinese collaborative research program MAECI-NSFC,
the Fond de la Recherche Scientifique (F.R.S-FNRS) and FWO under the ``Excellence of Science – EOS” in Belgium,
the Conselho Nacional de Desenvolvimento Cient\'ifico e Tecnol\`ogico in Brazil,
the Agencia Nacional de Investigacion y Desarrollo and ANID - Millennium Science Initiative Program - ICN2019\_044 in Chile,
the Charles University Research Centre and the Ministry of Education, Youth, and Sports in Czech Republic,
the Deutsche Forschungsgemeinschaft (DFG), the Helmholtz Association and its Recruitment Initiative, and the Cluster of Excellence PRISMA+ in Germany, the Joint Institute of Nuclear Research (JINR) and Lomonosov Moscow State University in Russia,
the joint Russian Science Foundation (RSF) and National Natural Science Foundation of China (NSFC) research program, the MOST and MOE in Taiwan,
the Chulalongkorn University and Suranaree University of Technology in Thailand, University of California at Irvine and the National Science Foundation in USA.



\begin{thebibliography}{99}


\bibitem{Rozwadowska:2021lll}
K.~Rozwadowska, F.~Vissani and E.~Cappellaro,
``On the rate of core collapse supernovae in the milky way,''
New Astron. \textbf{83}, 101498 (2021).

\bibitem{Ando:2004hc}
  S.~Ando and K.~Sato,
  ``Relic neutrino background from cosmological supernovae,''
  New J.\ Phys.\  {\bf 6}, 170 (2004).

\bibitem{Beacom:2010kk}
  J.~F.~Beacom,
  ``The Diffuse Supernova Neutrino Background,''
  Ann.\ Rev.\ Nucl.\ Part.\ Sci.\  {\bf 60}, 439 (2010).

\bibitem{Lunardini:2010ab}
  C.~Lunardini,
  ``Diffuse supernova neutrinos at underground laboratories,''
  Astropart.\ Phys.\  {\bf 79}, 49 (2016).

\bibitem{Malek:2002ns}
M.~Malek {\it et al.} [Super-Kamiokande Collaboration],
``Search for supernova relic neutrinos at SUPER-KAMIOKANDE,''
Phys.\ Rev.\ Lett.\  {\bf 90}, 061101 (2003).

\bibitem{Bays:2011si}
K.~Bays {\it et al.} [Super-Kamiokande Collaboration],
``Supernova Relic Neutrino Search at Super-Kamiokande,''
Phys.\ Rev.\ D {\bf 85}, 052007 (2012).


\bibitem{Super-Kamiokande:2021jaq}
K.~Abe \textit{et al.} [Super-Kamiokande],
``Diffuse supernova neutrino background search at Super-Kamiokande,''
Phys. Rev. D \textbf{104}, no.12, 122002 (2021).

\bibitem{Zhang:2013tua}
H.~Zhang {\it et al.} [Super-Kamiokande Collaboration],
``Supernova Relic Neutrino Search with Neutron Tagging at Super-Kamiokande-IV,''
Astropart.\ Phys.\  {\bf 60}, 41 (2015).


\bibitem{Beacom:2003nk}
J.~F.~Beacom and M.~R.~Vagins,
``GADZOOKS! Anti-neutrino spectroscopy with large water Cherenkov detectors,''
Phys.\ Rev.\ Lett.\  {\bf 93}, 171101 (2004).

\bibitem{Watanabe:2008ru}
H.~Watanabe {\it et al.} [Super-Kamiokande Collaboration],
``First Study of Neutron Tagging with a Water Cherenkov Detector,''
Astropart.\ Phys.\  {\bf 31}, 320 (2009).

\bibitem{Horiuchi:2008jz}
S.~Horiuchi, J.~F.~Beacom and E.~Dwek,
``The Diffuse Supernova Neutrino Background is detectable in Super-Kamiokande,''
Phys.\ Rev.\ D {\bf 79}, 083013 (2009).

\bibitem{Labarga:2018fgu}
L.~Labarga [Super-Kamiokande],
``The SuperK-gadolinium project,''
PoS \textbf{EPS-HEP2017}, 118 (2018).

\bibitem{KamLAND:2021gvi}
S.~Abe \textit{et al.} [KamLAND],
``Limits on Astrophysical Antineutrinos with the KamLAND Experiment,''
Astrophys. J. \textbf{925}, no.1, 14 (2022).

\bibitem{Collaboration:2011jza}
A.~Gando {\it et al.} [KamLAND Collaboration],
``A study of extraterrestrial antineutrino sources with the KamLAND detector,''
Astrophys.\ J.\  {\bf 745}, 193 (2012).

\bibitem{Borexino:2019wln}
M.~Agostini \textit{et al.} [Borexino],
``Search for low-energy neutrinos from astrophysical sources with Borexino,''
Astropart. Phys. \textbf{125}, 102509 (2021).

\bibitem{An:2015jdp}
  F.~An {\it et al.} [JUNO Collaboration],
  ``Neutrino Physics with JUNO,''
  J.\ Phys.\ G {\bf 43}, 030401 (2016).


\bibitem{Li:2013zyd}
Y.~F.~Li, J.~Cao, Y.~Wang and L.~Zhan,
``Unambiguous Determination of the Neutrino Mass Hierarchy Using Reactor Neutrinos,''
Phys. Rev. D \textbf{88}, 013008 (2013).

\bibitem{JUNO:2021vlw}
A.~Abusleme \textit{et al.} [JUNO],
``JUNO Physics and Detector,''
Prog. Part. Nucl. Phys. \textbf{123}, 103927 (2022).


\bibitem{JUNO:2022mxj}
A.~Abusleme \textit{et al.} [JUNO],
``Sub-percent Precision Measurement of Neutrino Oscillation Parameters with JUNO,''
Chin. Phys. C \textbf{46}, no.12, 123001 (2022),
[arXiv:2204.13249 [hep-ex]].

\bibitem{JUNO:2020hqc}
A.~Abusleme \textit{et al.} [JUNO],
``Feasibility and physics potential of detecting $^8$B solar neutrinos at JUNO,''
Chin. Phys. C \textbf{45}, no.2, 023004 (2021).


\bibitem{Han:2015roa}
R.~Han, Y.~F.~Li, L.~Zhan, W.~F.~McDonough, J.~Cao and L.~Ludhova,
``Potential of Geo-neutrino Measurements at JUNO,''
Chin. Phys. C \textbf{40}, no.3, 033003 (2016).

\bibitem{JUNO:2021tll}
A.~Abusleme \textit{et al.} [JUNO],
``JUNO sensitivity to low energy atmospheric neutrino spectra,''
Eur. Phys. J. C \textbf{81}, 10 (2021).


\bibitem{Lu:2016ipr}
J.~S.~Lu, Y.~F.~Li and S.~Zhou,
``Getting the most from the detection of Galactic supernova neutrinos in future large liquid-scintillator detectors,''
Phys. Rev. D \textbf{94}, no.2, 023006 (2016).


\bibitem{Priya:2017bmm}
  A.~Priya and C.~Lunardini,
  ``Diffuse neutrinos from luminous and dark supernovae: prospects for upcoming detectors at the $O$(10) kt scale,''
  JCAP {\bf 1711}, 031 (2017).



\bibitem{Kresse:2020nto}
D.~Kresse, T.~Ertl and H.~T.~Janka,
``Stellar Collapse Diversity and the Diffuse Supernova Neutrino Background,''
Astrophys. J. \textbf{909}, no.2, 169 (2021).


\bibitem{Horiuchi:2020jnc}
S.~Horiuchi, T.~Kinugawa, T.~Takiwaki, K.~Takahashi and K.~Kotake,
``Impact of binary interactions on the diffuse supernova neutrino background,''
Phys. Rev. D \textbf{103}, no.4, 043003 (2021).


\bibitem{ParticleDataGroup:2020ssz}
P.~A.~Zyla \textit{et al.} [Particle Data Group],
``Review of Particle Physics,''
PTEP \textbf{2020}, no.8, 083C01 (2020).

\bibitem{Salpeter:1955it}
E.~E.~Salpeter,
``The Luminosity function and stellar evolution,''
Astrophys. J. \textbf{121}, 161 (1955).


\bibitem{Hopkins:2006bw}
A.~M.~Hopkins and J.~F.~Beacom,
``On the normalisation of the cosmic star formation history,''
Astrophys. J. \textbf{651}, 142-154 (2006).


\bibitem{Strumia:2003zx}
A.~Strumia and F.~Vissani,
``Precise quasielastic neutrino/nucleon cross-section,''
Phys. Lett. B \textbf{564}, 42-54 (2003).


\bibitem{Cheng:2020aaw}
J.~Cheng, Y.~F.~Li, L.~J.~Wen and S.~Zhou,
``Neutral-current background induced by atmospheric neutrinos at large liquid-scintillator detectors: I. model predictions,''
Phys. Rev. D \textbf{103}, no.5, 053001 (2021).

\bibitem{Cheng:2020oko}
J.~Cheng, Y.~F.~Li, H.~Q.~Lu and L.~J.~Wen,
``Neutral-current background induced by atmospheric neutrinos at large liquid-scintillator detectors. II. Methodology for $in situ$ measurements,''
Phys. Rev. D \textbf{103}, no.5, 053002 (2021).

{
\bibitem{Mueller:2011nm}
T.~A.~Mueller, D.~Lhuillier, M.~Fallot, A.~Letourneau, S.~Cormon, M.~Fechner, L.~Giot, T.~Lasserre, J.~Martino and G.~Mention, \textit{et al.}
``Improved Predictions of Reactor Antineutrino Spectra,''
Phys. Rev. C \textbf{83}, 054615 (2011).
}

{
\bibitem{Huber:2011wv}
P.~Huber,
``On the determination of anti-neutrino spectra from nuclear reactors,''
Phys. Rev. C \textbf{84}, 024617 (2011)
[erratum: Phys. Rev. C \textbf{85}, 029901 (2012)].
}

\bibitem{Battistoni:2005pd}
G.~Battistoni, A.~Ferrari, T.~Montaruli and P.~R.~Sala,
``The atmospheric neutrino flux below 100-MeV: The FLUKA results,''
Astropart. Phys. \textbf{23}, 526-534 (2005).

\bibitem{Gaisser:1988ar}
T.~K.~Gaisser, T.~Stanev and G.~Barr,
``Cosmic Ray Neutrinos in the Atmosphere,''
Phys. Rev. D \textbf{38}, 85 (1988).

\bibitem{Honda:1990sx}
M.~Honda, K.~Kasahara, K.~Hidaka and S.~Midorikawa,
Phys. Lett. B \textbf{248}, 193-198 (1990).


\bibitem{Guo:2018sno}
W.~L.~Guo,
``Low energy neutrinos from stopped muons in the Earth,''
Phys. Rev. D \textbf{99}, no.7, 073007 (2019).

\bibitem{Hondahomepage}

The atmospheric neutrino fluxes at the JUNO site have been calcuated by the Honda group and made publicly available at http://www.icrr.u-tokyo.ac.jp/$\sim$mhonda


\bibitem{Abe:2009aa}
S.~Abe {\it et al.} [KamLAND Collaboration],
``Production of Radioactive Isotopes through Cosmic Muon Spallation in KamLAND,''
Phys.\ Rev.\ C {\bf 81}, 025807 (2010).

\bibitem{Bellini:2013pxa}
G.~Bellini {\it et al.} [Borexino Collaboration],
``Cosmogenic Backgrounds in Borexino at 3800 m water-equivalent depth,''
JCAP {\bf 1308}, 049 (2013).

\bibitem{Wang:2001fq}
Y.~F.~Wang, V.~Balic, G.~Gratta, A.~Fasso, S.~Roesler and A.~Ferrari,
``Predicting neutron production from cosmic ray muons,''
Phys. Rev. D \textbf{64}, 013012 (2001).

\bibitem{Honda:2015fha}
  M.~Honda, M.~Sajjad Athar, T.~Kajita, K.~Kasahara and S.~Midorikawa,
  ``Atmospheric neutrino flux calculation using the NRLMSISE-00 atmospheric model,''
  Phys.\ Rev.\ D {\bf 92}, no. 2, 023004 (2015).

\bibitem{Honda:2019ymh}
  M.~Honda, M.~Sajjad Athar, T.~Kajita, K.~Kasahara and S.~Midorikawa,
  ``Reduction of the uncertainty in the atmospheric neutrino flux prediction below 1 GeV using accurately measured atmospheric muon flux,''
  Phys.\ Rev.\ D {\bf 100}, no. 12, 123022 (2019).


\bibitem{Andreopoulos:2009rq}
C.~Andreopoulos, A.~Bell, D.~Bhattacharya, F.~Cavanna, J.~Dobson , S.~Dytman, H.~Gallagher, P.~Guzowski, R.~Hatcher, P.~Kehayias, A.~Meregaglia, D.~Naples, G.~Pearce, A.~Rubbia, M.~Whalley and T.~Yang,
``The GENIE Neutrino Monte Carlo Generator,''
Nucl. Instrum. Meth. A \textbf{614}, 87 (2010).
For further information, please refer to the website http://genie-mc.org/.

\bibitem{Golan:2012rfa}
T.~Golan, J.~Sobczyk and J.~Zmuda,
``NuWro: the Wroclaw Monte Carlo Generator of Neutrino Interactions,''
Nucl. Phys. B Proc. Suppl. \textbf{229-232}, 499-499 (2012).
Refer to the website https://nuwro.github.io/user-guide/ for further information.

\bibitem{Koning:2005ezu}
A.~Koning, S.~Hilaire and M.~Duijvestijn,
``TALYS: Comprehensive Nuclear Reaction Modeling,''
AIP Conf. Proc. \textbf{769}, no.1, 1154 (2005).


\bibitem{Mollenberg:2014pwa}
R.~M\"ollenberg, F.~von Feilitzsch, D.~Hellgartner, L.~Oberauer, M.~Tippmann, V.~Zimmer, J.~Winter and M.~Wurm,
``Detecting the Diffuse Supernova Neutrino Background with LENA,''
Phys. Rev. D \textbf{91}, no.3, 032005 (2015)
[arXiv:1409.2240 [astro-ph.IM]].


\bibitem{Hocker:2007ht}
A.~Hocker, P.~Speckmayer, J.~Stelzer, J.~Therhaag, E.~von Toerne, H.~Voss, M.~Backes, T.~Carli, O.~Cohen and A.~Christov, \textit{et al.}
``TMVA - Toolkit for Multivariate Data Analysis,''
[arXiv:physics/0703039 [physics.data-an]].

\bibitem{Pedregosa:2012toh}
F.~Pedregosa, G.~Varoquaux, A.~Gramfort, V.~Michel, B.~Thirion, O.~Grisel, M.~Blondel, G.~Louppe, P.~Prettenhofer and R.~Weiss, \textit{et al.}
``Scikit-learn: Machine Learning in Python,''
J. Machine Learning Res. \textbf{12}, 2825-2830 (2011)


\bibitem{JUNO:2020xtj}
A.~Abusleme \textit{et al.} [JUNO],
``Calibration Strategy of the JUNO Experiment,''
JHEP \textbf{03}, 004 (2021)
[arXiv:2011.06405 [physics.ins-det]].


\bibitem{Feldman:1997qc}
G.~J.~Feldman and R.~D.~Cousins,
``A Unified approach to the classical statistical analysis of small signals,''
Phys. Rev. D \textbf{57}, 3873-3889 (1998).


\bibitem{Li:2022myd}
Y.~F.~Li, M.~Vagins and M.~Wurm,
``Prospects for the Detection of the Diffuse Supernova Neutrino Background with the Experiments SK-Gd and JUNO,''
Universe \textbf{8}, no.3, 181 (2022)
[arXiv:2201.12920 [astro-ph.HE]].

%

%

%
%

%
%
%
%
%



\bibitem{Moller:2018kpn}
  K.~M{\o}ller, A.~M.~Suliga, I.~Tamborra and P.~B.~Denton,
  ``Measuring the supernova unknowns at the next-generation neutrino telescopes through the diffuse neutrino background,''
  JCAP {\bf 1805}, 066 (2018).


\bibitem{deGouvea:2020eqq}
A.~De Gouv\^ea, I.~Martinez-Soler, Y.~F.~Perez-Gonzalez and M.~Sen,
``Fundamental physics with the diffuse supernova background neutrinos,''
Phys. Rev. D \textbf{102}, 123012 (2020).


\bibitem{Abe:2018uyc}
  K.~Abe {\it et al.} [Hyper-Kamiokande Collaboration],
  ``Hyper-Kamiokande Design Report,''
  arXiv:1805.04163 [physics.ins-det].


\bibitem{DUNE:2020ypp}
B.~Abi \textit{et al.} [DUNE],
``Deep Underground Neutrino Experiment (DUNE), Far Detector Technical Design Report, Volume II: DUNE Physics,''
[arXiv:2002.03005 [hep-ex]].


\bibitem{Wurm:2011zn}
  M.~Wurm {\it et al.} [LENA Collaboration],
  ``The next-generation liquid-scintillator neutrino observatory LENA,''
  Astropart.\ Phys.\  {\bf 35}, 685 (2012).


\bibitem{Askins:2019oqj}
M.~Askins \textit{et al.} [Theia],
``THEIA: an advanced optical neutrino detector,''
Eur. Phys. J. C \textbf{80}, no.5, 416 (2020).


\end{thebibliography}
\end{document}